\documentclass[titlepage,12pt]{article}

\usepackage{authblk}
\usepackage[margin=1in]{geometry}

\usepackage{algorithm}
\usepackage{algpseudocode}
\usepackage{amsbsy, amsfonts, amsmath, amssymb}

\usepackage{soul} 
\usepackage{graphicx}
\usepackage[colorlinks=true, citecolor=blue]{hyperref}
\usepackage{mathrsfs}
\usepackage{bm}
\usepackage{multirow}
\usepackage{booktabs, threeparttable}
\usepackage{setspace}
\usepackage{tikz}
\usepackage{caption} 
\usepackage{diagbox}
\usepackage{subfigure}

\usepackage{natbib}

\usepackage{physics} 

\usepackage{color}
\usepackage{xcolor}

\newcommand{\blue}[1]{\textcolor{blue}{#1}}

\usepackage[inline]{enumitem}

\allowdisplaybreaks
\newcommand{\adj}{\bm{A}}
\newcommand{\wei}{\bm{W}}
\newcommand{\din}{d^{\, \rm (in)}}
\newcommand{\dout}{d^{\, \rm (out)}}
\newcommand{\stin}{s^{\, \rm (in)}}
\newcommand{\stout}{s^{\, \rm (out)}}
\newcommand{\bc}{\bm{c}}
\newcommand{\bB}{\bm{B}}

\newcommand{\normoutw}{\tilde{w}^{\, \rm (out)}}
\newcommand{\norminw}{\tilde{w}^{\, \rm (in)}}

\graphicspath{{./}{./image/}}

\begin{document}

\title{Regional and Sectoral Structures and Their Dynamics of 
Chinese Economy: A Network Perspective from Multi-Regional 
Input-Output Tables}

\author[1]{Tao Wang}
\author[1]{Shiying Xiao}
\author[2]{Jun Yan}
\author[3,$\ast$]{Panpan Zhang}
\affil[1]{School of Statistics, Shanxi University of Finance and
Economics, Taiyuan 030006, China}
\affil[2]{Department of Statistics, University of Connecticut,
Storrs, CT 06269, USA}
\affil[3]{Department of Biostatistics, Epidemiology and Informatics, 
University of Pennsylvania, Philadelphia, PA 19104, USA}
\affil[$\ast$]{Corresponding author. Email:
\href{mailto:panpan.zhang@pennmedicine.upenn.edu}
{panpan.zhang@pennmedicine.upenn.edu}}

\maketitle

\doublespacing

\begin{abstract}
A multi-regional input-output table (MRIOT) containing the
transactions among the region-sectors in an economy defines a
weighted and directed network. Using network analysis tools, we
analyze the regional and sectoral structure of the Chinese economy
and their temporal dynamics from 2007 to 2012
via the MRIOTs of China. Global analyses are done with network
topology measures. Growth-driving province-sector clusters are
identified with community detection methods. Influential
province-sectors are ranked by weighted PageRank scores. The 
results revealed a few interesting and telling  insights. The level of
inter-province-sector activities increased with the rapid growth of
the national economy, but not as fast as that of intra-province economic
activities. Regional community structures were deeply associated with
geographical factors. The community heterogeneity across the regions
was high and the regional fragmentation
increased during the study period. Quantified metrics assessing the
relative importance of the province-sectors in the national economy
echo the national and regional economic development policies to a
certain extent.

\end{abstract}

\doublespacing

\section{Introduction}
\label{sec:intr}

The rapid growth of the Chinese economy over the last three decades 
has drastically elevated its importance in the global economy.
The annual growth rate of gross domestic product (GDP) during 1990--2010
was 10.4\% \citep{IMF2020gdp}.
The growth in China was a driving force for the recovery of the world
from the financial crisis in 2008 \citep{lin2011china}. As the Chinese
economy matured, the growth slowed down to 6.74\% over 2015--2019, but it was
still much higher than that of the world economy, 2.82\%, during the same period
\citep{TWB2020growth}. In 2019, China contributed 16.34\% to the global GDP,
second only to the United States and almost tripling the contribution by Japan
which  ranked the third \citep{TWB2020gdp}. China has become an integrated part
of the global economy. As a top trader, China accounted for 10.14\% 
of the global
imports \citep{TWB2020imports} and 10.61\% of the global exports in 2019 
\citep{TWB2020exports}.  Behind the growth, there have been dramatic structural
changes such as urbanization and industrialization \citep{fan2003structural,
  chen2011structural}. The regional and sectoral structures of the
Chinese economy are heavily affected by internal
government's policies such as the Great Western Development Strategy 
\citep{jia2020place} or external factors such as the World Trade
Organization accession in 2001 \citep{chow2003impact}
and the 2008 global financial crisis \citep{yuan2010impact}.

Given its size and impact, the Chinese economy is central to important regional
and sectoral structure issues in economic development theory and 
practice.
The disparities in sectoral structure and economic growth at the
province level in China are high and have been increasing
\citep[e.g.,][]{fan2011china, li2011economic, lee2012regional}.
Liberalized and globalized industries are mostly aggregated in the
coastal regions while low technology, resource-based, and
protected industries are widely dispersed in the inland regions
\citep[e.g.,][]{he2012regional}.
Emerging industries are more likely to enter the regions that are 
globalized,
economically liberalized and fiscally healthy 
\citep{he2018regional}. On one hand, the distribution of value-added 
across regions has been flattened
due to the expansion of interregional trade 
\citep{meng2017measuring}.
On the other hand, growing inter-regional competition and local 
protection have jointly led
to various inter-regional trade barriers and severe regional 
fragmentation
\citep{young2000razor, poncet2005fragmented}.
Understanding the regional and sectoral structures of the Chinese
economy is critical for economic development and resource efficiency
not only in China but also the entire world.

Multi-regional input-output tables (MRIOTs) are the most prevalent tool
for studying the inter-dependencies among the sectors from different 
regions in an economy.
An MRIOT records the transactions among the sectors within multiple
regions \citep[e.g.,][]{moses1955stability, leontief1963multiregional}.
A few MRIOTs at the global level have been available with bilateral trade
information for a large number of countries annually for several
decades \citep{tukker2013global}. MRIOTs at the province level
within China, however, are not readily available. Based on 
the survey-based input-output tables at the province level released 
by the government, Chinese MRIOTs have been compiled for 2007 and 
2012 \citep{liu2012mriot, liu2018mriot}. The tables and their
extensions have been used in the analyses of the impact of government
infrastructural plans \citep{ji2019energy}, provincial and 
sector-level material footprints \citep{liang2017structural, 
jiang2019provincial}, and cross-country sectoral price comparison 
\citep{fujikawa2002input}, among others.

An MRIOT inherently defines a weighted directed network which
facilitates analyses with statistical methods for networks.
The world input-output tables (WIOTs) \citep{timmer2015illustrated} are 
MRIOTs at the global level. Traditional tools for input-output table 
analysis include multipliers, linkages, and structural paths
\citep{defourny1984structural, feser2000national}, which measure the 
impact from each region-sector in the 
table. In contrast, network analysis tools
enable not only the measures at the region-sector level such as 
centrality, but also the natural investigation on local clustering, 
community detection, and
backbone extraction as well as global network features such as assortativity and
clustering coefficient \citep{leonidov2019dynamical, xu2019input}. 
With a sequence of input-output
tables over different years, the dynamic changes of network features can be
investigated, which are of great value in structural and regional analyses
\citep{cerina2015world, del2017trends, amador2017networks}.
For the Chinese MRIOTs, in part due to their limited availability, no network
analysis has been done to study the regional and sectoral structure of
the Chinese economy.

Our contributions are two-fold. 
First, to the best of our knowledge, this is the first comprehensive
network analysis of the MRIOTs of China to study regional and 
sectoral structure of the Chinese economy. Through the analysis, 
we have evidently observed a clear pattern of increased regional 
fragmentation from 2007 to 2012. Some of the 
research outcomes have not been reported in the literature, and 
may not be straightforwardly uncovered via traditional input-output
analysis tools. Our second contribution is the application of several
novel network measures specifically developed for weighted and
directed networks to analyze the Chinese MRIOTs. The MRIOTs lead
to networks that are both weighted and directed. If weight and/or
direction are disregarded like in some existing analyses, the
features of the networks such as assortativity and centrality
are not precisely summarized. The new
network measures help correct the misleading results and inaccurate
inference driven from the classical unweighted versions.

The rest of the manuscript is organized as follows. In
Section~\ref{sec:mriot}, we give a brief introduction of the 
compiled MRIOTs
in China, and demonstrate the network analysis setup. The specific
network analysis methods are presented in Section~\ref{sec:meth},
followed by the applications to the MRIOTs in Section~\ref{sec:res}.
Finally, we address some concluding remarks and follow-up
discussions in Section~\ref{sec:dis}.

\section{MRIOTs of China}
\label{sec:mriot}

The MRIOTs of China are available for the year of 2007 \citep{liu2012mriot},
2010 \citep{liu2014mriot}, and 2012 \citep{liu2018mriot}. The databases were
jointly developed by the Institute of Geographic Sciences and
the Natural Resources Research of the Chinese Academy of Sciences, 
and the National Bureau of Statistics of China. The entries of
inter-province-sector economic transactions were obtained by 
applying the
gravity model \citep{bergstrand1985gravity, sargento2007empirical} to the
input-output tables reported by all the participating provinces.
Table~\ref{tab:mriot} shows the fundamental structure of MRIOT.
We focus on the 2007 and 2012 tables in the present study because 
the compilation of the 2010 table was based on the 2007 table in 
addition to the input-output tables of the 17 provinces
rather than direct data collection and investigation \citep{mi2018multi},
which might cause measurement errors and bias. An 
MRIOT consists of 
four parts:
\begin{enumerate*}[label = (\Roman*)]
\item intermediate flow matrix,
\item final use,
\item imports, and
\item value added.
\end{enumerate*}
The intermediate flow matrix records the
economic exchanges among the sectors from different
provinces, reflecting their intricate economic relations (e.g.,
supply and demand) as well as their interdependence and mutual
constraints.

\begin{table}[tbp]
	\footnotesize
	\renewcommand*{\arraystretch}{1.5}
	\centering
	\caption{Fundamental tructure of an MRIOT.}
	\label{tab:mriot}
	\resizebox{\textwidth}{!}{
		\begin{tabular}{c|c|c|c|c|c|c|c|c|c|c|c}
			\hline
			\multicolumn{3}{c|}{\multirow{3}{*}{}}
			& \multicolumn{7}{c|}{Intermediate use}
			& \multirow{3}{*}{\begin{tabular}[c]{@{}c@{}}
					Final \\ use \end{tabular}}
			& \multirow{3}{*}{\begin{tabular}[c]{@{}c@{}}
					Total \\ output \end{tabular}} \\
			\cline{4-10}
			\multicolumn{3}{c|}{} & \multicolumn{3}{c|}{region 01}
			& $\cdots$ & \multicolumn{3}{c|}{region 30} & & \\
			\cline{4-10}
			\multicolumn{3}{c|}{} & sector 01 & $\cdots$ & sector 30
			& $\cdots$ & sector 01 & $\cdots$ & sector 30 & & \\
			\hline
			\multirow{7}{*}{
				\rotatebox[origin=c]{90}{Intermediate input}}
			& \multirow{3}{*}{\rotatebox[origin=c]{90}{region 01}}
			& sector 01
			& \multicolumn{7}{c|}{\multirow{7}{*}
				{\uppercase\expandafter{\romannumeral 1}}}
			& {\multirow{7}{*}
				{\uppercase\expandafter{\romannumeral 2}}}
			& {\multirow{7}{*}{}} \\
			\cline{3-3}
			& & $\vdots$ & \multicolumn{7}{c|}{} & &  \\
			\cline{3-3}
			& & sector 30 & \multicolumn{7}{c|}{} & & \\
			\cline{2-3}
			& $\vdots$ & $\vdots$ & \multicolumn{7}{c|}{} & & \\
			\cline{2-3}
			& \multirow{3}{*}{\rotatebox[origin=c]{90}{region 30}}
			& sector 01
			& \multicolumn{7}{c|}{} & & \\
			\cline{3-3}
			& & $\vdots$ & \multicolumn{7}{c|}{} & & \\
			\cline{3-3}
			& & sector 30 & \multicolumn{7}{c|}{} & & \\
			\hline
			\multicolumn{3}{c|}{Imports}
			& \multicolumn{7}{c|}{
				\uppercase\expandafter{\romannumeral 3}}
			& \multicolumn{2}{c}{\multirow{2}{*}{}} \\
			\cline{1-10}
			\multicolumn{3}{c|}{Value added}
			& \multicolumn{7}{c|}{
				\uppercase\expandafter{\romannumeral 4}}
			& \multicolumn{2}{c}{\multirow{2}{*}{}} \\
			\cline{1-10}
			\multicolumn{3}{c|}{Total input} &
			\multicolumn{7}{c|}{}
			& \multicolumn{2}{c}{} \\
			\hline
	\end{tabular}}
\end{table}

The data in the MRIOTs were pre-processed to prepare for the 
analyses. 
The 2007 table covered 30 provincial units with each containing 30 
sectors. The 2012 table,
however, covered 31 provincial units due to the debut of Tibet and 42 sectors
that were further divided from the 30 sectors in the 2007 table.
For the purpose of comparison over time, we only included the 30 
provinces that appeared
in both tables and aggregated the 42 sectors in 2012 to the 30 
sectors in 2007. Table~\ref{tab:sec} lists the codes with detailed descriptions 
for the 30
sectors. The monetary units for both tables
were set to be 10,000 Chinese Yuan (CNY). To adjust for inflation,
we converted the entries in the 2012 table to 2007 CNY
using the GDP price deflator \citep{TWB2020inflation}.

\begin{table}[tbp]
  \centering
  \small
	\caption{Description of the sectors in the MRIOTs}
	\label{tab:sec}
	\begin{tabular}{p{0.8cm}<{\centering} p{6.7cm}
			p{0.8cm}<{\centering}
			p{6.7cm}}
		\toprule
		Code & \multicolumn{1}{c}{Sector} & Code &
		\multicolumn{1}{c}{Sector} \\
		\midrule
		01 & Agriculture, forestry, animal husbandry and fishery
		& 16 & General and specialist machinery \\
		02 & Coal mining and processing
		& 17 & Transport equipment \\
		03 & Petroleum and gas extracting
		& 18 & Electrical equipment \\
		04 & Metals mining/processing
		& 19 & Electronic equipment \\
		05 & Nonmetal mining/processing
		& 20 & Instrument and meter \\
		06 & Food processing and tobaccos
		& 21 & Other manufacturing \\
		07 & Textiles
		& 22 & Electricity and heat production and supply \\
		08 & Clothing, leather, fur, etc.
		& 23 & Gas and water production and supply \\
		09 & Wood processing and furnishing
		& 24 & Construction \\
		10 & Paper making, printing, stationery, etc.
		& 25 & Transport and storage \\
		11 & Petroleum refining, coking, etc.
		& 26 & Wholesale and retail \\
		12 & Chemical industry
		& 27 & Hotel and restaurant \\
		13 & Nonmetal products
		& 28 & Leasing and commercial services \\
		14 & Metallurgy
		& 29 & Scientific research \\
		15 & Metal products
		& 30 & Other services \\
		\bottomrule
	\end{tabular}
\end{table}

\begin{figure}[tbp]
  \centering
  \includegraphics[width=.99\textwidth]{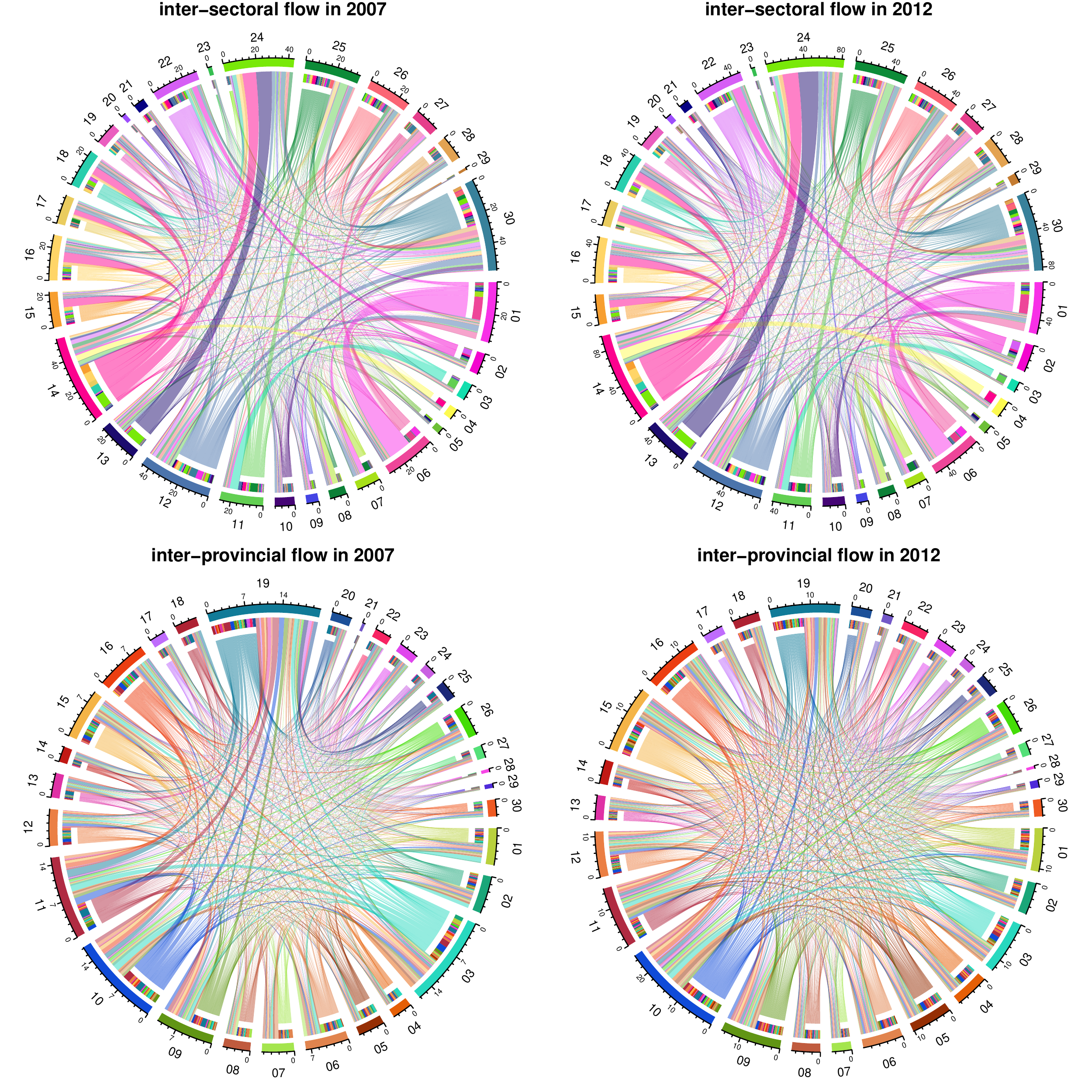}
  \caption{Intermediate flows (across multiple regions) in the 
  MRIONs from 2007 and 2012. The 
  top two panels are for sectors (aggregated with respect to 
  provinces), where the codes are referred to Table~\ref{tab:sec}. 
  The bottom two panels are for provinces (aggregated with respect 
  to sectors), where the codes are referred to 
  Appendix~\ref{sec:provcode}. The bandwidth of chord (connecting 
  the arcs) is proportional to the size of economic flow. Long arcs 
  indicate large outputs. The unit of economic flow is 100 billion 
  CNY.}
  \label{fig:chord}
\end{figure}

We constructed the multi-regional input-output networks (MRIONs)
based on the MRIOTs. In an MRION, each vertex represents a sector
within a province; each directed edge represents the existence
of transaction from the source province-sector to the target
province-sector, with weight representing the multiplier of the
transaction 10,000 CNY. Therefore, the
MRIONs are weighted and directed. The
number of vertices in each MRION is 900. The link
densities are respectively 0.7685 in 2007 and 0.9212 in 2012,
suggesting that the vertices in the MRIONs are densely connected.
The top two panels of Figure~\ref{fig:chord} show chord 
visualizations of the MRIONs aggregated according to
sectors for 2007 and 2012 with self-loops removed.
Each of the outer arcs with a distinct color represents a sector, 
with arc length
representing the sum of the inflows and outflows. A chord from one arc to
another represents the transaction from the corresponding sector to 
the other. Its width is proportional to the volume of the 
transaction, while its color remains the same as the
color of the source sector. For both years, the main suppliers are 
``metallurgy'' (14), ``chemical industry'' (12), ``other services'' (30), and 
``agriculture, forestry, animal husbandry and fishery'' (01). They 
supply 
a large portion of the
intermediate products or services that are needed by other sectors.
The most notable receivers are ``construction'' (24) and ``other
services'' (30). These two sectors may have strong pulling effects on 
the whole
economy. Especially in ``construction'' (24), the proportion of inflows
in its inter-sectoral transaction exceeds 90\%.
One notable change from 2007 to 2012 is the share of the transactions
associated with sector ``scientific research'' (29), which is quadrupled
from 0.24\% to 0.97\%.

In addition, we provide the chord visualizations of the MRIONs 
aggregated by provinces (with self-loops removed as well) for 2007 
and 2012, shown in the bottom two panels in Figure~\ref{fig:chord}. 
The arcs and chords are defined analogously as the top ones. The 
main suppliers in 
2007 were Hebei (03), Guangdong (19) and Jiangsu (10), but Hebei 
(03)  was replaced with Shandong (15) in 2012, indicating that the 
production capacity of Shandong (15) got strongly (from 2007 to 
2012). For both years, the most notable receivers were Jiangsu 
(10), Zhejiang (11) and Guangdong (19). These three provinces have 
promoted a large number of inter-provincial trade exchanges with the 
others across the nation. In fact, the majority of the provinces 
presenting high inter-provincial trade amount were from the 
coastal region. In spite of the substantial drops in the
proportions of inter-provincial trade in Guangdong (19) and Zhejiang 
(11) from 2007 to 2012, they were still top~5 over the nation and 
remained the driving forces contributing to the multi-regional 
economy in China. More quantitative assessment calls for detailed 
network analyses.

\section{Methods}
\label{sec:meth}

Our approach to investigating the MRIOTs of China are network-based
analytics. Let $G(V,E)$ denote a directed network that consists of a
set of vertices $V$ and a set of edges $E$. By convention, each vertex
$i \in V$ represents a data point. Given a pair of vertices
$i, j \in V$, if there is a directed edge from $i$ to $j$, then we 
have $e_{ij} \in E$. Vice versa, the existence of an edge $e_{ij} 
\in E$ suggests a
(directed) link from $i$ to $j$. One of the most popular
ways of displaying a network structure is adjacency matrix.
For a network with $n = |V|$ vertices, its adjacency matrix is
denoted by $\adj := (a_{ij})_{n \times n}$ with
$a_{ij} = 1$ if $e_{ij} \in E$ and $a_{ij} = 0$ otherwise.
For a weighted and directed network, the weighted counterpart is
denoted by $\wei := (w_{ij})_{n \times n}$, where $w_{ij}$
represents the weight of edge $e_{ij}$. The weighted adjacency
matrix $\wei$ is equivalent to $\adj$ if $w_{ij} = 1$ for all
$e_{ij} \in E$.

\subsection{Degree and strength distributions}
\label{subsec:degree}

In a directed network, the degree of a vertex $i$ (denoted
$d_i$) is comprised of in-degree (denoted by $\din_i$) and
out-degree (denoted by $\dout_i$), which are, respectively,
the number of edges pointing into and emanating out of vertex~$i$.
To account for edge weight, we define the in-strength and 
out-strength of vertex $i$ as $\stin_i := \sum_{j \in V} w_{ji}$ and
$\stout_i := \sum_{j \in V} w_{ij}$, respectively. The strength of 
vertex~$i$, $s_i$,
is the sum of its in-strength and out-strength. In traditional 
network analyses, degrees and strengths are used to show the
importance of the vertices in a network \citep{newman2010networks}.

The degree distribution is the probability distribution $\pi(\cdot)$ of the
vertex degrees over the entire network; that is, $\pi(k)$ is the
probability of a vertex having degree $k \in \{ 0, 1, 2, \ldots\}$.
The degree distribution plays an important role in theoretical and
applied network analyses. In a completely random
network~\citep{erdos1959on}, the degree distribution is Poisson, 
whereas the tail of the degree
distribution of a scale-free network~\citep{barabasi1999emergence}
follows a power law. \citet{pennock2002winners} pointed out that most
real networks fall between these two
extreme classes. It is evident that the degree distributions of
economic networks are likely to exhibit power-law
patterns~\citep{gabaix1999zipf, kaplow2008pareto}.
The goodness-of-fit of power-law tails can be tested based on the
Kolmogorov--Smirnov statistic~\citep{clauset2009power} with
p-values obtained from bootstrapping.

In a weighted network, the strength distribution, which is based on
the vertex strengths, usually better captures the network structure
than the degree distribution. While the degree distribution is
always discrete, the strength distribution can be either discrete or
continuous, depending on the characteristic of weight. As the MRIONs
are weighted and directed, we conducted analogous analyses on the
strength, in-strength and out-strength distributions and made
comparisons with their degree counterparts.

\subsection{Assortativity}
\label{subsec:assort}

Assortativity (or assortative mixing)
refers to the tendency that the vertices in a network are
connected according to a pair of (vertex-specific) features
\citep{newman2002assortative}.
It is a measure of homophily among the vertices based on two
given features. A commonly used assortativity
measures is the degree-degree correlation
\citep{newman2002assortative,vanderhofstad2014degree},
which is analogous to Pearson correlation coefficient.
Its value is between $-1$ and $1$, with a positive (negative) value
indicating that high-degree vertices are more likely to be connected 
with high-degree (low-degree) vertices. This measure is
amenable to directed networks \citep{newman2003mixing, foster2010edge}.
See \citet{noldus2015assortativity} for a comprehensive survey.

Since the MRIONs are not only directed but also weighted, we
adopted a class of assortativity measures proposed
by \citet{yuan2020assortativity} to incorporate edge weight and 
direction. Let
$(\alpha,\beta) \in \{{\rm in}, {\rm out}\}$ index the type of
strength. The assortativity based on $\alpha$- towards $\beta$-type
strength is
\begin{equation}
  \label{eq:assort}
  \displaystyle
  \rho_{\alpha, \beta}(G) =
  \frac{
    \sum_{i, j \in V} w_{ij}
    \left[\left(s_i^{(\alpha)} - \bar{s}_{\rm sou}^{\, (\alpha)}
      \right)
      \left(s_j^{(\beta)} - \bar{s}_{\rm tar}^{\, (\beta)}
      \right)\right]
  }
  {W \sigma_{\rm sou}^{(\alpha)} \sigma_{\rm tar}^{(\beta)}},
\end{equation}
where $W := \sum_{i,j \in V} w_{ij}$ is the total weight,
$s_{i}^{(\alpha)}$ is the $\alpha$-type strength of source vertex $i$,
$s_{j}^{(\beta)}$ is the $\beta$-type strength of target vertex $j$,
\begin{equation*}
  \bar{s}_{\rm sou}^{\, (\alpha)} = \frac{\sum_{i,j \in V} w_{ij}
    s_i^{\, (\alpha)}}{W}
  \mbox{ and }
    \bar{s}_{\rm tar}^{\, (\beta)} = \frac{\sum_{i,j \in V} w_{ij}
    s_j^{\, (\beta)}}{W}
\end{equation*}
are, respectively, the weighted mean of the
$\alpha$-type strength of the source vertices and
$\beta$-type strength of the target vertices,
and
\begin{equation*}
  \sigma_{\rm sou}^{(\alpha)} = \sqrt{\frac{\sum_{i,k \in V}
      w_{ik} \left(s_i^{(\alpha)} - \bar{s}_{\rm sou}^{\,
          (\alpha)}
      \right)^2}{W}}
  \mbox{ and }
  \sigma_{\rm tar}^{(\beta)} = \sqrt{\frac{\sum_{k, j \in V}
      w_{kj} \left(s_j^{(\beta)} - \bar{s}_{\rm tar}^{\,
          (\beta)}
      \right)^2}{W}}
\end{equation*}
are the associated weighted standard deviations.
A positive (negative) $\rho_{\alpha,\beta}(G)$ suggests
assortative-mixing (disassortative-mixing), and zero assortativity
indicates no obvious pattern of assortative- or
disassortative-mixing.

For a network like MRION, the weighted adjacency matrix can be 
decomposed into one comprised of diagonal blocks only and another 
comprised of off-diagonal blocks. The former contains the 
information of economic transactions within each province (called 
intra-province), while the latter records the exchanges across 
multiple provinces (called inter-province). The proposed 
assortativity measure can be applied to the decomposed adjacency 
matrices to investigate the correlation structures at the intra- and 
inter-province levels. 

\subsection{Clustering coefficient}
\label{subsec:clustering}

Clustering coefficient is a measure quantifying the tendency
that the vertices in a network are clustered together, usually
characterized by a high density of connections among
them \citep{opsahl2009clustering}. Clustering coefficient is also
known as transitivity coefficient in the
literature~\citep{newman2002random}. Classical clustering
coefficient was proposed for undirected and unweighted
networks~\citep{watts1998collective}, and later were extended to
weighted and directed networks~\citep{grindrod2002range,
barrat2004architecture, onnela2005intensity, zhang2005general,
fagiolo2007clustering,
clemente2018directed}.

In the present study, we adopted the weighted and directed clustering
coefficients developed by~\citet{clemente2018directed}. The local 
clustering coefficient of vertex~$i$ in an unweighted and undirected 
network $G(V, E)$ is the ratio of the number of links connecting the 
neighbors 
of $i$ (i.e., $\{j \in V : e_{ij} 
\in E\}$) to the maximum possible value. When edges have weights, the
weighted adjacency matrix $\wei$ plays an important role.  
Self-loops are removed prior to the computation
since they do not practically contribute to the network clustering 
property.
The clustering coefficient can be concisely expressed via matrix notations:
\begin{equation}
\label{eq:clust}
C^{\, \rm tot}_i = \frac{\left[\left(\wei +
\wei^{\top}\right)\left(\adj +
\adj^{\top}\right)^2 \right]_{ii}}{2\left[s_i (d_i - 1) - \left(\adj
\wei + \wei\adj\right)_{ii}\right]},
\end{equation}
where $\adj^{\top}$ is the transpose of $\adj$, and $B_{ii}$
is the $(i,i)$th element of matrix $B$.

The superscript
distinguishes it from the four kinds of distinct local clustering
coefficients induced from four types of directed
triangles \citep{fagiolo2007clustering, clemente2018directed}.
Namely, they are in-, out-, mid- and cyc-clustering coefficients. 
When computing a specific local clustering coefficient, the 
denominator needs to be updated to the number of corresponding 
triplets. For instance, the local in-clustering coefficient of $i$ 
is the number of triangles such that the neighbors (say, $j$ 
and $k$) both link towards $i$ alongside with an edge (in either
direction) connecting $j$ and $k$ out of the number of triplets with 
both $j$ and $k$ generating directed edges to 
$i$ (disregarding whether or not $j$ and $k$ are connected). All of 
the other local clustering coefficients are defined in an analogous 
manner. The local
out-clustering coefficient of $i$ is the proportion of triangles which have
two edges from $i$ pointing to $j$ and~$k$ and an edge
linking $j$ and $k$ in either direction. The local
mid-clustering coefficient of $i$ considers the proportion of the 
triangles in which
$i$ is a middleman: neighbor $j$ (or $k$)
either has a direct link to neighbor $k$ (or $j$) or forms
a directed path $j \rightarrow i \rightarrow k$ (or $k \rightarrow i
\rightarrow j$). The local cyc-clustering coefficient of $i$ only
counts the triangles of which the directed edges form a cycle. See
Figure~\ref{fig:triangles} in Appendix~\ref{sec:triangle} for
graphic illustrations and the formulae therein for practical
computation. Accordingly, there are five kinds of global clustering
coefficients on the network base, obtained by averaging the
associated local clustering coefficients over all the vertices.

Similar to assortativity, any kind of clustering coefficients 
introduced in this section can be applied to the decomposed 
adjacency matrices of MRION to uncover the clustering properties at
the intra- and intro-provincial levels.

\subsection{Community detection}
\label{subsec:community}

Community detection aims to group the entities with similar 
characteristics in a
network to the same community.
The entities in the same community are densely linked, while those
from different communities are loosely linked. There are two major
classes of community detection methods, model-based
\citep{snijders1997estimation, handcock2007model} and
metric-based \citep{girvan2002community, ouyang2020clique}. In this
study, we used a metric-based method for community detection in
MRIONs. Interested readers are referred to \citet{goldenberg2010survey}
for a comprehensive survey for community detection techniques.

Specifically, we exploited the modularity maximization algorithm
proposed by~\citet{newman2006modularity}. An objective
function called modularity is defined to measure the quality of
clustering strategies, and is then maximized.
The underlying principle of modularity maximization is that the
number of links among the vertices within a community is
significantly more than expected at random (based on the
Erd\"{o}s--R\'{e}nyi model which is generally used as the null
model), while the counterpart across different communities is 
significantly less.
Newman's algorithm is built upon recursive bi-partitioning. For a
weighted and undirected network $G(V, E)$, its modularity
matrix $\bB := (b_{ij})_{n \times n}$ is defined as
\[ b_{ij} := w_{ij} - \frac{s_i s_j}{W},
\]
where $W$ is identical to that defined in
Section~\ref{subsec:assort}. The term $s_i s_j / W$ is
interpreted as the expected weight of the edges connecting $i$
and $j$ if all the edges are randomly placed among the vertices in
the network.
Let $\bc := (c_i)_{i = 1}^{n}$ denote a
clustering strategy. A bi-partitioning algorithm admits two
clusters, so $c_i$'s take value $1$ or $-1$ representing distinct
membership. Given $\bc$, we define a modularity score as
\[
  Q = \frac{1}{W} \sum_{i,j} b_{ij}
  I(c_i = c_j),
\]
where $I(\cdot)$ is the indicator function.

The expression of $Q$ can be regarded as a
reward-penalty system. Given $c_i = c_j$, the value of $Q$ increases 
if $b_{ij} > 0$, but decreases if $b_{ij} < 0$. Besides,
the larger $b_{ij}$ is (given $c_i = c_j$), the more reward is 
granted. Subsequent
bi-partitioning continues within each resulting community until no
more partition in any existing community leads to an increase in
modularity score.

For large networks, parsimonious
algorithms~\citep{clauset2004finding, ng2001spectral, ouyang2020clique}
are needed to solve the optimization problem. We
adopted the greedy algorithm developed by \citet{clauset2004finding}.

\subsection{Centrality}
\label{sec:centrality}

The centrality of each vertex measures its relative importance
in a network. Vertices with high centrality scores
altogether form the main frame of the network.
There are various ways of defining centrality depending on
practical needs and interpretations, such as degree
centrality \citep{barrat2004architecture},
closeness and betweenness \citep{newman2001scientific}, and
eigenvector centrality \citep{bonacich1987power}.
We considered a measure extended from eigenvector
centrality, namely PageRank \citep[PR,][]{brin1998anatomy},
that was originally used for ranking websites by Google.

We propose an extension of classical PR for weighted and directed
networks. This extension is different from the existing ones that
are specifically designed for analyzing citation
networks \citep{xing2004weighted, ding2011applying}. We define
the PR centrality of vertex~$i \in V$ as
\begin{equation}
\label{eq:pagerank}
P_i = \gamma \sum_{j \in V} \left(\theta \, \frac{w_{ji}}{\stout_j} 
+ (1 - \theta) \frac{a_{ji}}{\dout_j} \right) P_j +
\frac{(1 -
  \gamma)\beta_i}{\sum_{i \in V} \beta_i},
\qquad i = 1, \ldots, n,
\end{equation}
where $\theta \in [0, 1]$ is a tuning parameter indicating the 
proportion of edge weight (versus edge number) accounted in PR, 
$\gamma$ is a damping factor that prevents the algorithm from
getting stuck in sinking vertices (those without outgoing edges),
and $\beta_i$ is a prior measure (usually
independent of network structure) of the relative
importance of vertex~$i$. When there is no information available for
$\gamma$, it takes value $0.85$ as suggested by \citet{page1998pagerank}.
In spite of the prior information specified
by $\beta_i$'s, Equation~\eqref{eq:pagerank} suggests that a vertex
receives a high PR score (with $\theta = 1$) if
\begin{enumerate*}[label = (\roman*)]
	\item it receives a large number of incoming edges from the
	others in the network;
	\item the weights of the incoming edges linking to it are large;
	\item the senders themselves have high PR scores.
\end{enumerate*}

When there is no prior information about $\beta_i$'s,
they can be set the same, in which case the
second term in Equation~\eqref{eq:pagerank} is simplified to
$(1 - \gamma)/n$. A standard method to solve
Equation~\eqref{eq:pagerank} is power iteration, but the convergence of this
algorithm may be slow for large-scale networks. A remedy is to
utilize the stochastic process theory and convert the problem to
finding the stable distribution of an underlying Markov Chain
\citep{berkhin2005survey}. The investigations of the crucial 
properties of the proposed PR 
measure will be reported elsewhere.

\subsection{Backbone}
\label{sec:backbone}

The backbone of a network is the fundamental but essential structure of a
network \citep{xu2019input}. Non-essential links, which act like 
noise in a
large network, can be removed without affecting the backbone. 
Extracting the
backbone of a massive and dense network like MRION is critical, as
hundreds of edges with minimal weights would overwhelm
the analysis. Proper removal of non-essential edges helps succinctly 
characterize a complex network system, and meanwhile enhance 
computation speed. There
have been a few promising backbone extraction methods,
such as the disparity filter method \citep{serrano2009extracting,
zhang2013skeleton}, the locally adaptive network sparsification
algorithm \citep{foti2011nonparametric}, and two classes of
node-based filtering approaches \citep{ghalmane2020extracting}.
We used the disparity filter method which has been applied to
the analysis of WIOTs by \citet{xu2019input}.

The rational of the disparity filter method is as follows. Consider the
normalized weights of the $d \in \mathbb{Z}^{+}$ edges of a vertex. 
Under the null hypothesis that
the normalized weights are generated from a uniform random assignment, they can
be regarded as obtained by dividing the unit interval by $(d-1)$ randomly placed
points. The lengths of the subintervals, which represent the
normalized weights, have density function
\[
  p(x; d) = (d - 1)(1 - x)^{d - 2}, \quad x \in (0, 1).
\]
An overly large normalized weight relative to this distribution means that
the corresponding edge is unlikely to be from the uniform random assignment,
which supports the corresponding
edge to be part of the backbone. This idea can be formulated as obtaining
the p-value of each normalized weight. Define $\tilde w_{ij} = 
w_{ij} / s_i$. The
p-value of $\tilde w_{ij}$ is
\[
  \delta_{ij}(\tilde w_{ij}; d_i) = \int_{\tilde w_{ij}}^1 p(x; d_i) \dd x.
\]
The backbone with level $\alpha\in (0, 1)$ is obtained by retaining only those
edges whose p-values are less than $\alpha$.

For a directed network, normalized out- and
in-strength of each edge can be defined similarly as
$\normoutw_{ij} = w_{ij} / \stout_i$ and
$\norminw_{ij} = w_{ij} / \stin_j$ for all $i, j \in V$. The
corresponding p-values
are $\delta_{ij}(\normoutw_{ij}; \dout_i)$ and
$\delta_{ij}(\norminw_{ij}; \din_j)$, respectively.
For backbone with level $\alpha \in (0, 1)$, edge $e_{ij}$ is
preserved if at least one of the two p-values is less than~$\alpha$.

\begin{figure}[tbp]
	\centering
	\includegraphics[width=\textwidth]{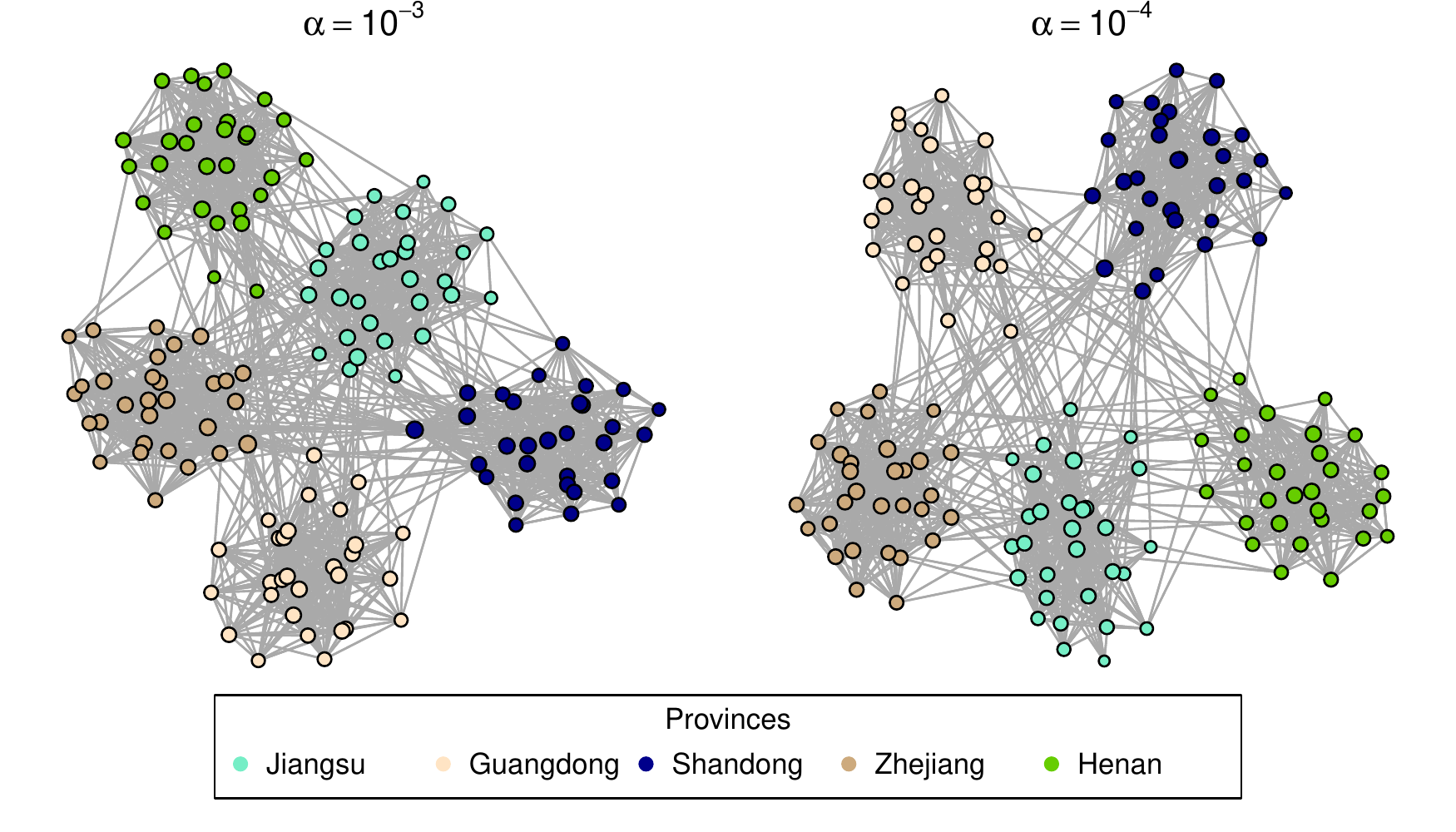}
	\caption{Examples of the sub-MRIONs comprised of the sectors 
	from the provinces with top~5 regional GDP in 2012. The 
	self-loops are not presented. The significance levels of 
	$\alpha$ are respectively $10^{-3}$ (left panel) and $10^{-4}$ 
	(right panel). }
    \label{fig:backbone2012}
\end{figure}

In some rare cases of $\dout_i = 1$, $\din_i = 1$ or  $\dout_i
= \din_i = 1$, special treatments may be needed, depending on the
specific features of the networks as well as the practical
interpretation of network heterogeneity. These rare cases do not 
occur in our study. In Figure~\ref{fig:backbone2012}, we present the 
sub-networks of the MRION in 2012 consisting of the sectors only 
from top~5 regional GDP provinces with significance level $\alpha 
\in \{10^{-3}, 10^{-4}\}$. Though the sub-networks remain dense, 
they relatively better reflect the basic structure of the MRION, and 
furthermore, suggest province-based market fragmentation as well as 
community structure, which are consistent with some of the results 
shown in Section~\ref{sec:res}.

\section{Results}
\label{sec:res}

We apply the methods in Section~\ref{sec:meth} to the 2007 and
2012 MRIONs of China, and present the corresponding results.
The interpretations of the analysis results are given from 
both statistical and economic perspectives.

\subsection{Degree and strength distributions}
\label{sec:res_deg}

\begin{figure}[tbp]
	\centering
	\includegraphics[width=\textwidth]{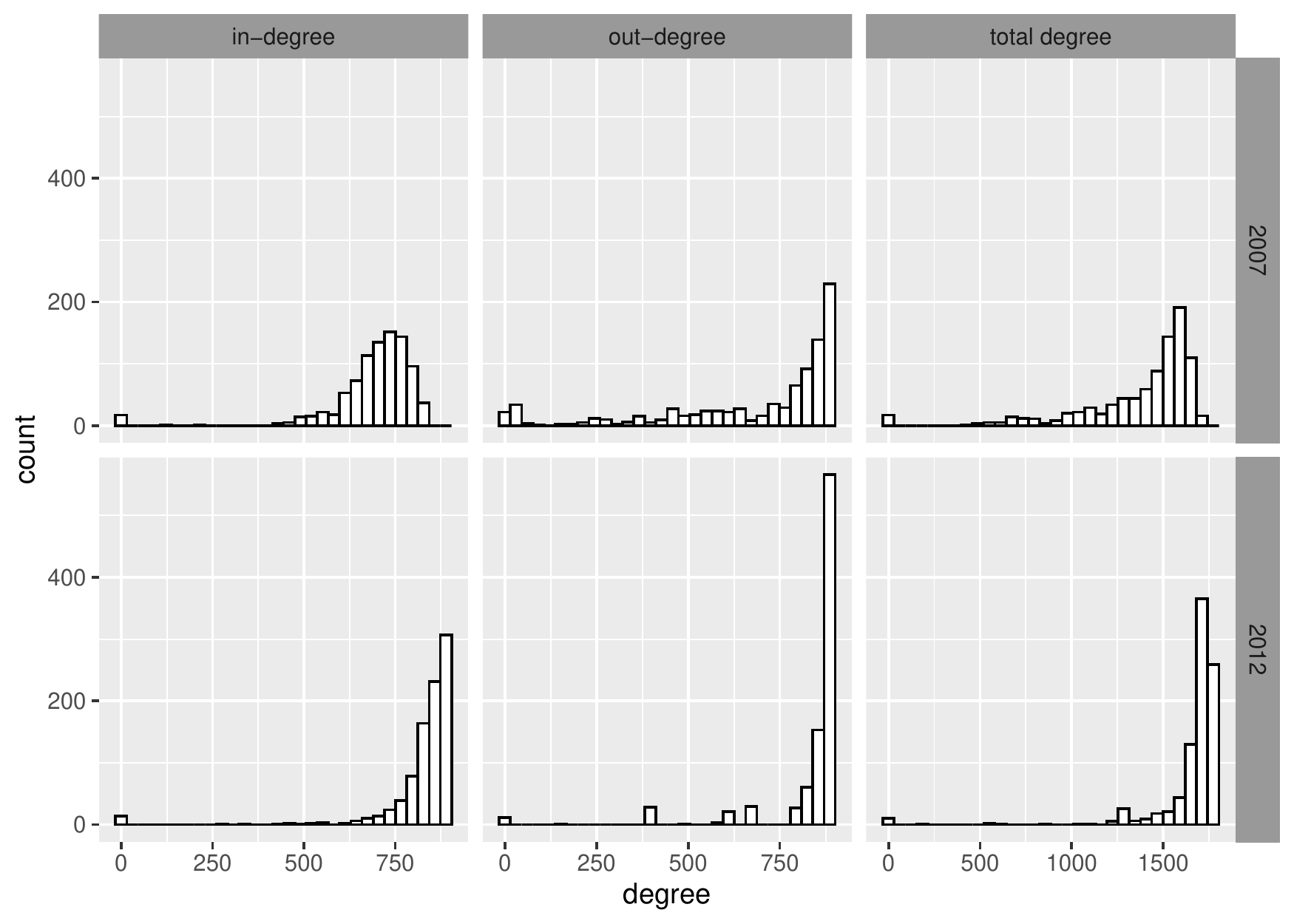}
	\caption{In-, out- and total-degree distributions of the MRIONs 
	in 2007 and 2012.}
	\label{fig:degdist}
\end{figure}

Figure~\ref{fig:degdist} shows the histograms for the 
in-, out- and total-degree distributions of the MRIONs in
2007 and 2012. These degree distributions appear to share two features. First,
there is a strictly positive probability of zero degree, which corresponds to
province-sectors with no edges (i.e., isolated vertices). For 
example, the sectors ``coal mining and processing'' (02), ``metals 
mining/processing'' (04) and ``nonmetal mining/processing'' (05) in 
Shanghai are singletons with neither inbound nor outbound links.
Second, the nonzero degrees are close to the maximum degree and 
skewed to the left, more significantly reflected in out-degrees 
than in-degrees. This is a result of
heavily connected province-sectors. From 2007 to 2012, all three degree
distributions shifted to the right with more left skewness, 
suggesting the increase in the number of links among the 
province-sectors during this period in China.

\begin{figure}[tbp]
	\centering
	\includegraphics[width=\textwidth]{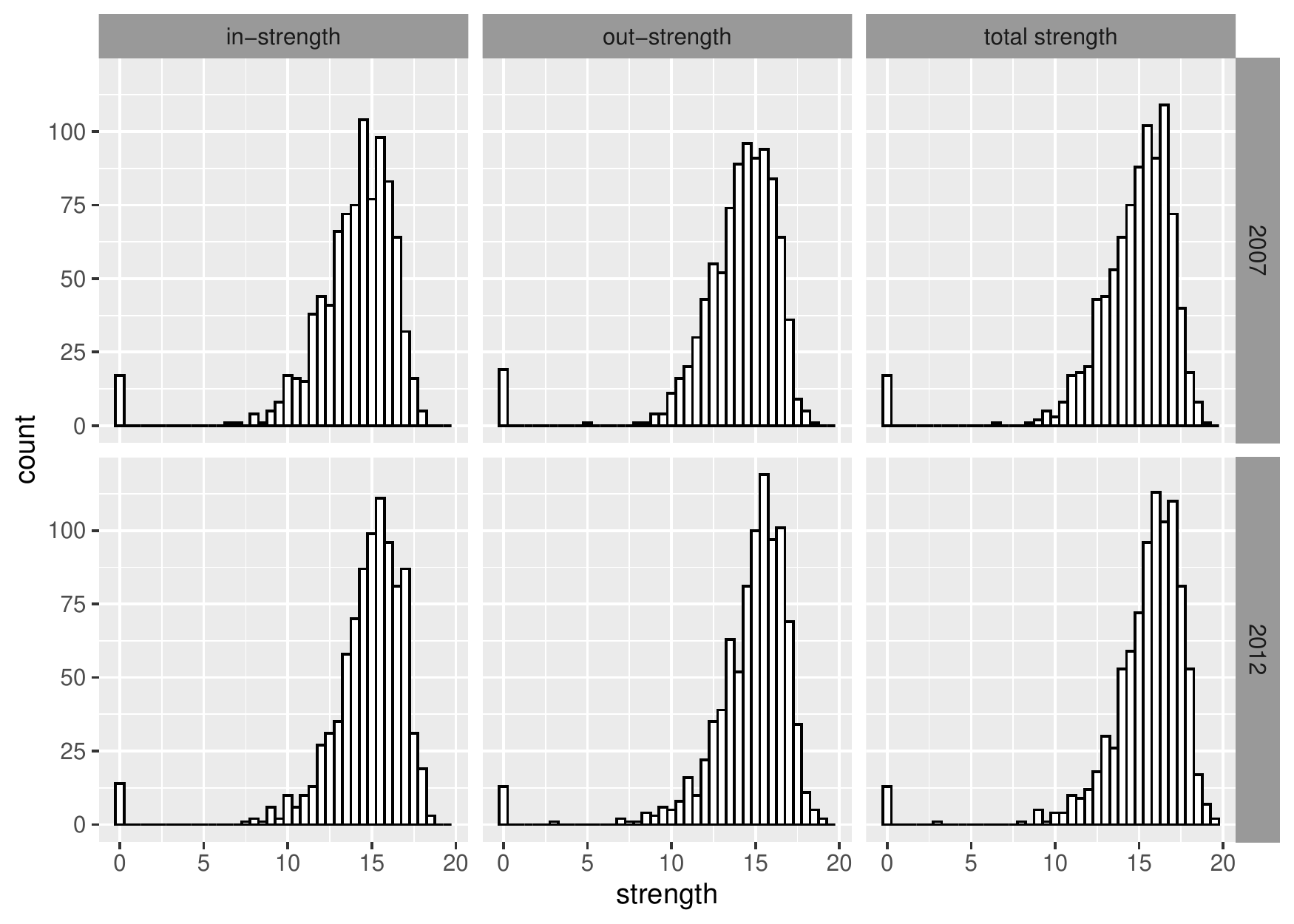}
	\caption{In-, out- and total-strength distributions
	(after logarithmic transformation) of the MRIONs
	in 2007 and 2012.}
	\label{fig:strdist}
\end{figure}


More information about the magnitude of the economic transactions is provided in
the strength distributions on the log scale shown in Figure~\ref{fig:strdist}.
The strength distributions are mixtures of a point mass at zero and a positive
continuous distribution. Such distributions are often used to model
zero-inflated non-negative continuous data through the mechanism of two-part
models or hurdle models \citep{liu2019statistical}.
The mass at zero is inherited from the zero degrees from the degree
distributions. The positive strengths on the log scale are skewed to the left.
On the original scale, however, the positive strengths are skewed to the right.
Similar to the degree distributions, the strength distributions all shifted to
the right from 2007 to 2012 with most of the quartiles more than doubled,
reflecting an expansion of the
economic transactions among the province-sectors during this period.

\begin{figure}[tbp]
  \centering
  \includegraphics[width=\textwidth]{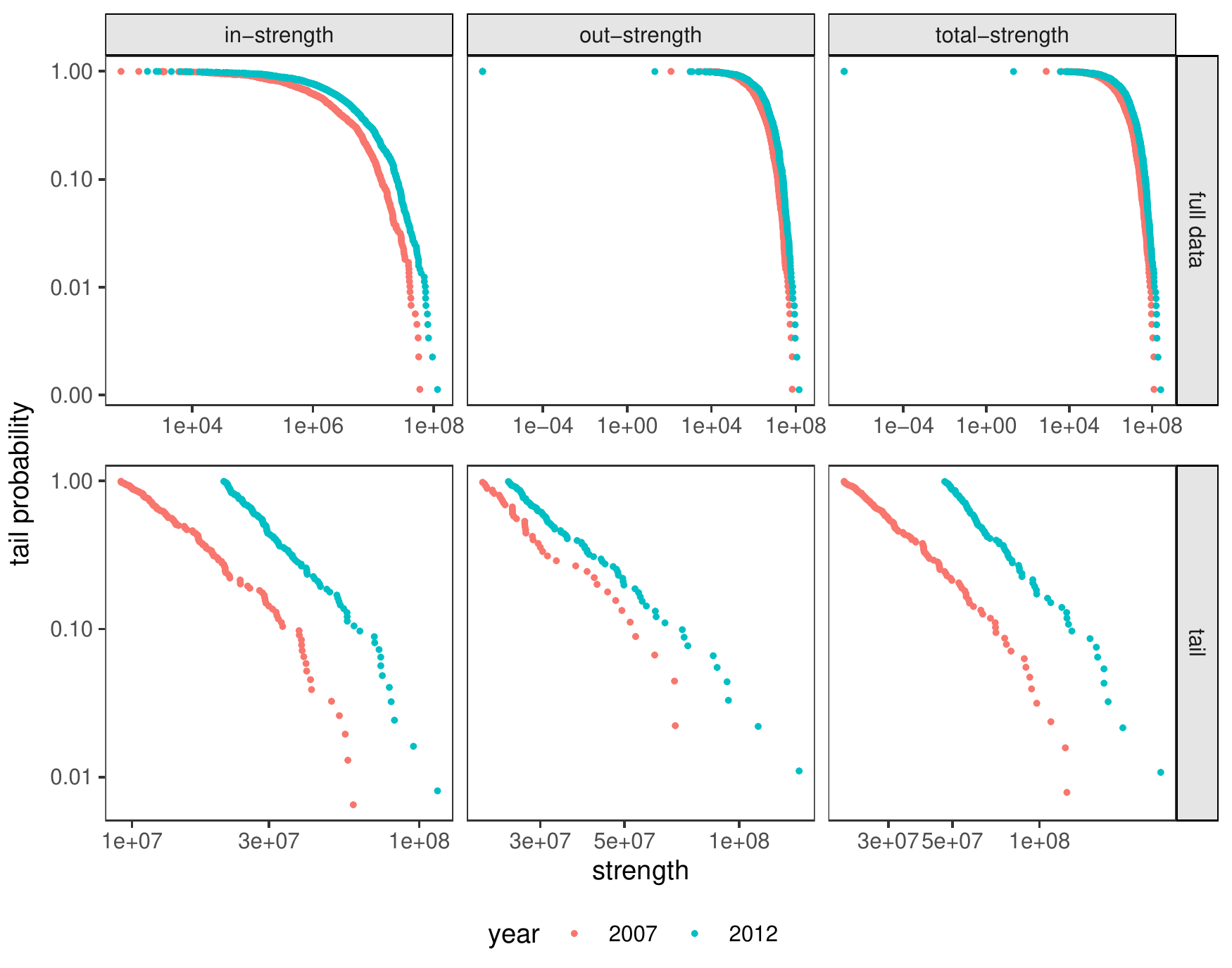}
  \caption{Tail distributions of in-, out- and total-strength of
  	the MRIONs in 2007 and 2012 (bottom two panels).
  	Their correspondingly zoomed tails after the estimated thresholds
  	are given in the top two panels.}
  \label{fig:strtaildist}
\end{figure}

The tail of the strength distribution is of great importance, 
especially in extreme value theory, as it characterizes the features 
of the distribution far way from the mean, indicating the relative 
probability of the occurrence of some ``unusual'' events, i.e., 
extensively large strength sectors. Specifically, we are interested 
in a particular class of heavy tail distributions---power laws, 
which, as mentioned in Section~\ref{subsec:degree}, have been 
observed in a variety of economic networks.
Figure~\ref{fig:strtaildist} shows the empirical survival curves of 
the three strength distributions in 2007 and 2012 with both axis on 
the log scale. Empirical
distributions of such shapes appear to be typical for MRIONs
\citep[e.g.,][]{xu2019input}. The tails of the distributions show 
plausible
linear patterns (on the log-scale) which are the characteristics of 
power laws. To 
verify, we performed goodness-of-fit tests for power law tails
\citep{clauset2009power}. Table~\ref{tab:plstat} summarizes the estimated
thresholds and exponent parameters as well as the p-values of the 
power law tails
beyond the estimated thresholds obtained from bootstrapping. The power law
provides adequate fit to all three strength distributions in 2012 and the
out-strength distribution in 2007; it is rejected at significance level 0.01 for
the in-strength and total-strength distribution in 2007.
This is consistent with the lower panels of
Figure~\ref{fig:strtaildist}, which show the empirical conditional survival
curves beyond the estimated thresholds. For the out-strength, the smaller
exponent parameter estimate in 2012 than in 2007 indicates heavier tails in the
magnitude of the extremely large transactions in 2012 than in 2007.

\begin{table}[tbp]
	\centering
	\caption{Estimated parameters for power-law tails for the in-,
	out- and total-strength distributions of the MRIONs in 2007 and 2012
	and p-values of the goodness-of-fit test; the unit of threshold
	is $1$ million CNY.}
	\small \setlength{\tabcolsep}{3.7pt}
	\label{tab:plstat}
	\begin{tabular}{ccccccc}
		\toprule
		 & \multicolumn{3}{c}{2007} & \multicolumn{3}{c}{2012}
		\\ \cmidrule(lr){2-4} \cmidrule{5-7}
		 & In-strength & Out-strength & Total-strength
		 & In-strength & Out-strength & Total-strength
		\\ \midrule
		Threshold & $9.11$ & $20.91$ & $20.99$ & $20.83$ & 
		$24.41$ & $46.96$
		\\ 
		Exponent & $2.65$ & $3.56$ & $2.84$ & $3.19$ & $3.29$ & $3.33$
		\\[1ex]
		p-value & $0.00$ & $0.27$ & $0.01$ & $0.18$ & $0.46$ & $0.42$
		\\ \bottomrule
	\end{tabular}
\end{table}

\subsection{Assortativity}
\label{sec:res_assort}

Table~\ref{tab:assort} summarizes a collection of assortativity coefficients for
the MRIONs of 2007 and 2012. For each year, the assortativity coefficients were
computed for both directed (four types) links and undirected links;
intra-province links, inter-province links, and nationwide links; unweighted links and
weighted links.

\begin{table}[tbp]
	\centering
	\caption{Five kinds of assortativity
		coefficients (including ``total'' that does not account for 
		edge direction) of the (national, intra-province and 
		inter-province) MRIONs in 2007 and 2012, where ``UW'' and 
		``W'' respectively represents the unweighted and weighted 
		versions of the assortativity measures.}
	\label{tab:assort}
	\small \setlength{\tabcolsep}{3pt}
	\begin{tabular}{l rrrrrr rrrrrr}
		\toprule
		\multirow{4}{*}{Type} & \multicolumn{6}{c}{2007}
		& \multicolumn{6}{c}{2012} \\ \cmidrule{2-7} 
		\cmidrule(lr){8-13}
		& \multicolumn{2}{c}{National} & 
		\multicolumn{2}{c}{Intra-prov.}
		& \multicolumn{2}{c}{Inter-prov.}
		& \multicolumn{2}{c}{National}
		& \multicolumn{2}{c}{Intra-prov.} & 
		\multicolumn{2}{c}{Inter-prov.} \\
		\cmidrule(lr){2-3} \cmidrule(lr){4-5} \cmidrule(lr){6-7}
		\cmidrule(lr){8-9} \cmidrule(lr){10-11} \cmidrule(lr){12-13}
		& UW & W & UW & W & UW & W & UW & W & UW & W & UW & W \\
		\midrule
		in-in & $-0.010$ & $0.501$ & $0.258$ & $0.610$ & $-0.033$ & 
		$0.023$ & $-0.003$ & 
		$0.573$ & $0.209$ & $0.649$ & $-0.008$ & $-0.022$ \\
		in-out & $-0.001$ & $0.447$ & $0.124$ & $0.549$ & $-0.006$ & 
		$-0.024$ & $-0.001$ & 
		$0.516$ & $0.109$ & $0.594$ & $-0.002$ & $-0.034$ \\
		out-in & $-0.123$ & $0.493$ & $0.007$ & $0.599$ & $-0.136$ & 
		$0.117$ & $-0.110$ & 
		$0.563$ & $0.010$ & $0.635$ & $-0.111$ & $0.130$ \\
		out-out & $-0.024$ & $0.474$ & $0.070$ & $0.572$ & $-0.028$ 
		& $0.047$ & $0.015$ & 
		$0.536$ & $0.095$ & $0.618$ & $0.016$ & $-0.001$ \\
		total & $-0.070$ & $0.418$ & $0.140$ & $0.544$ & $-0.080$ & 
		$0.046$ & $-0.030$ & 
		$0.457$ & $0.158$ & $0.561$ & $-0.031$ & $0.028$ \\
		\bottomrule
	\end{tabular}
\end{table}

Our first observation is that the unweighted assortativity coefficients
\citep{newman2002assortative, foster2010edge} are not informative 
for characterizing the
MRIONs. The unweighted versions are similar to the weighted versions only for
inter-province links, which are
all close to zero. For nationwide links and intra-province links, the weighted and
unweighted assortativity coefficients of all five types, including four directed
and one undirected, are notably different. The maximum magnitude
of the unweighted version is only $0.123$ (out-in in 2007), which is much lower
than the magnitudes of the weighted versions in the range of 0.4--0.6. The
two versions have completely different signs for all five types of nationwide
links in 2007. The unweighted versions suggest that there is a negligible
pattern of disassortative mixing, while the weighted versions suggest
assortative mixing. The weighted versions are more consistent with intuition.
Some existing analyses of the WIOTs without weight also reported 
``close to zero''
assortativity coefficients \citep[e.g.,][]{cerina2015world}, which need to be
revisited by using the weighted definition 
\citep{yuan2020assortativity}.

Based on the results from the weighted definitions, the assortativity
coefficients for nationwide links of all kinds have magnitude 
from~$0.4$
to~$0.6$. These are moderately strong assortative mixing. Take the out-in
assortativity for nationwide links as an example. The coefficient of 
$0.493$ in 2007 and $0.563$ in 2012 suggest that province-sectors 
with large inputs are likely to take high
transaction volumes from the others with high outputs in the network.
Decomposing the weighted adjacency matrix helps separate the 
contributions respectively from intra-province and inter-province 
links. 
All five assortativity coefficients for intra-province links are much
greater than those for inter-province links, which are close to zero. That is,
economic transactions among the sectors from the same province are 
extensively close, with high transaction volumes; in contrast,
sectors across multiple regions are relatively loosely connected, 
and the majority of the transaction volumes represented by existing 
link weights are extremely small. The differences between intra- and
inter-province assortativity coefficients support the well-known 
regional fragmentation \citep{poncet2005fragmented}.

The assortativity coefficients of all types for nationwide links and
intra-province links increased from 2007 to 2012, while those for inter-province
links remained close to zero. The increases in nationwide links are therefore
attributed to the increase in intra-province links, suggesting an increase in
the degree of provincial segmentation. One possible explanation, for example, is
that the economic stimulus plan after the 2008 financial crisis stimulated the
construction industry most, which propagated to upstream metal and non-metal
mining/processing sectors within each province. Despite the increased
inter-provincial transactions, there is no clear assortative pattern 
among these transactions in contrast to the intra-provincial transactions.

\begin{figure}[tbp]
	\centering
	\includegraphics[width=\textwidth]{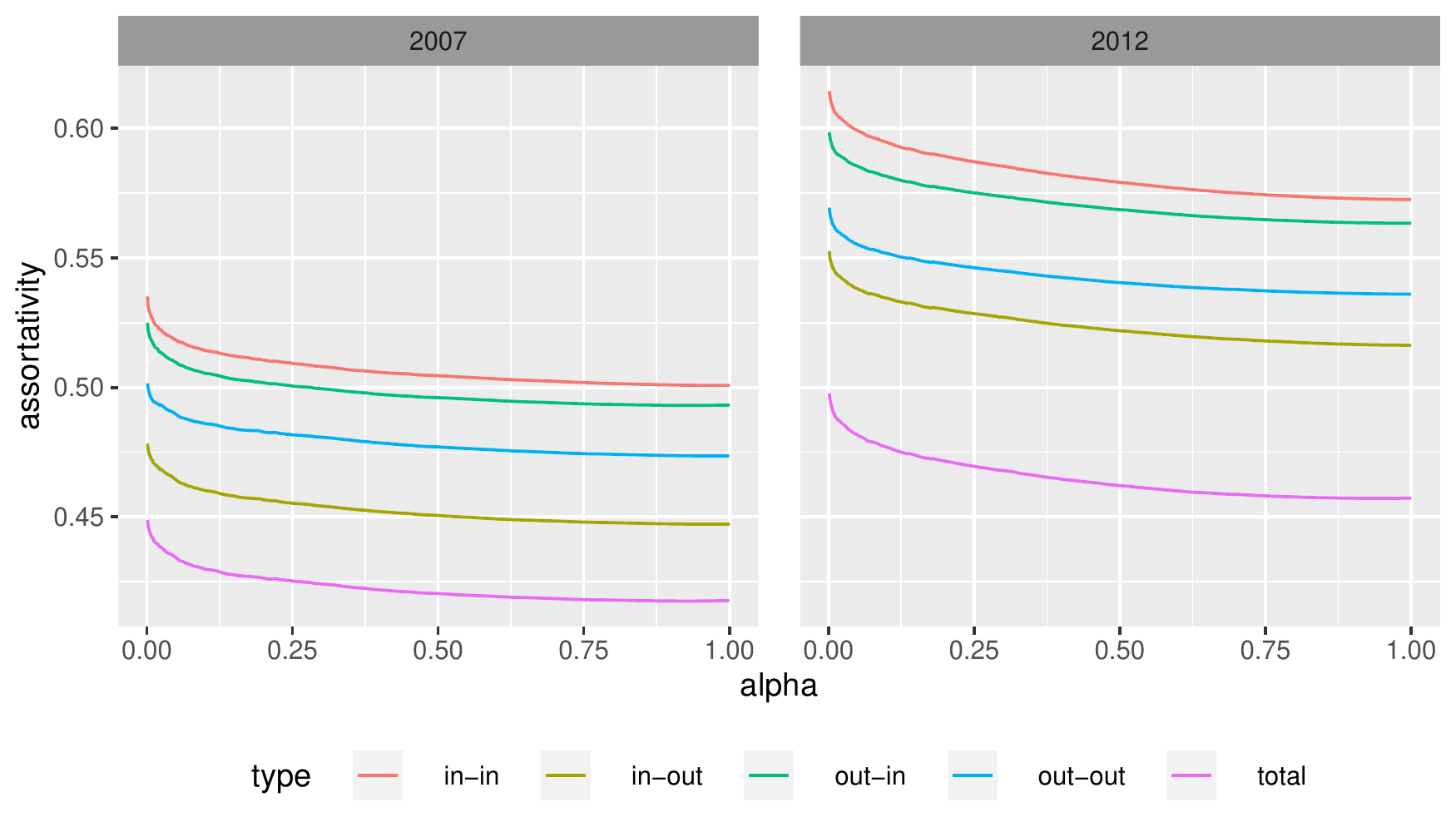}
	\caption{Five kinds of assortativity coefficients
		(national level) at a sequence of significance levels, 
		$\alpha = \{0.001, 0.002, \ldots, 0.999\}$ in 2007 
		and 2012.}
	\label{fig:assort_backbone}
\end{figure}

The assortativity coefficient provides a platform to demonstrate the
effectiveness of backbone. Figure~\ref{fig:assort_backbone} shows the national
assortativity coefficients for the backbones of the MRIONs in 2007 and
2012 for a sequence of significance levels $\alpha = \{0.001, 0.002, 
\ldots, 0.999\}$. For each year, the value of each type of assortativity
coefficient increases, but only slightly, with the decrease of $\alpha$ 
due to the removal of the non-essential edges. The removed edges 
are supposed to impose limited impact on the overall structure of 
the network. As a result, the magnitude of change in each 
assortativity coefficient is small. Further, for 
each $\alpha$, all kinds of the assortativity coefficients in 2012 
are greater than their counterparts in 2007, which is consistent 
with the results from Table~\ref{tab:assort}. Therefore, backbone is a 
parsimonious and powerful tool for uncovering the fundamental and 
essential properties of a network, especially for the large-scale 
networks that are likely to cause computational expensiveness.

\subsection{Clustering coefficients}
\label{sec:res_clust}

\begin{table}[tbp]
	\centering
	\caption{Total, cycle-, middleman-, in- and out-clustering
		coefficients of the MRIONs at national, intra-province and 
		inter-province levels in 2007 and 2012.}
	\label{tab:clustcoeff}
	\begin{tabular}{lcccccc}
		\toprule
		\multirow{3}{*}{Type} & \multicolumn{3}{c}{2007}
		& \multicolumn{3}{c}{2012} \\ \cmidrule{2-4} 
		\cmidrule(lr){5-7}
		& National & Intra-prov. & Inter-prov.
		& National & Intra-prov. & Inter-prov. \\
		\midrule
		Total & 0.874 & 0.942 & 0.855 & 0.968 & 0.969 & 0.937 \\ 
		Cycle & 0.828 & 0.933 & 0.815 & 0.960 & 0.962 & 0.933 \\ 
		Middleman & 0.927 & 0.952 & 0.888 & 0.975 & 0.975 & 0.942 \\ 
		In & 0.914 & 0.949 & 0.881 & 0.968 & 0.971 & 0.935 \\ 
		Out & 0.843 & 0.939 & 0.819 & 0.966 & 0.967 & 0.939 \\ 
		\bottomrule
	\end{tabular}
\end{table}

Clustering coefficients were computed with and without edge 
direction for three types
of links, nationwide, intra-province, and inter-province, in the 
2007 and 2012 MRIONs;
see Table~\ref{tab:clustcoeff}. All the clustering coefficients have large values 
(close to $1$), providing stronger evidence than simple link 
densities for immense connectivity of the MRIONs. The larger values of 2012 
suggest a higher tendency that the province-sectors would cluster 
together in terms of forming triangles. Decomposing the nationwide links to
intra-province and inter-province links reveals that the nationwide increase
from 2007 to 2012 was mainly due to the increase in 
inter-province components. For example, the nationwide 
cycle-clustering coefficient increased from 0.828 to 0.960; the intra-province
coefficient 0.933 in 2007 was quite high, leaving not much room to increase;
the inter-province coefficient increased from 0.815 in 2007 to 0.933 in 2012.
The emergence of more transactions across the inter-province sectors may be
attributed to the goverment's strategies and policies such as the Great 
Western Development Strategy. The increase in inter-province transaction is not
in contradition to its small proportion in the ovreall magnitude, so it did not
affect the manifestation of regional fragmentation in China.

Among the four types of directed clustering coefficients, the cycle- and 
out-clustering coefficients have increased more 
notably than the others. In a cyclic-triangle 
connection, each province-sector is an upstream as well as a 
downstream of its neighbors. A higher value of cycle-clustering 
coefficient indicates a higher proportion of triangular (supply and 
demand) chains formed by the province-sectors. In an out-triangle connection, a
province-sector is always the upstream to its neighbors. A higher
value of out-clustering coefficient suggests
an increased proportion transactions among the downstream sectors.

\subsection{Community detection}
\label{sec:res_community}

\begin{figure}[tbp]
  \centering
  \includegraphics[width=\textwidth]{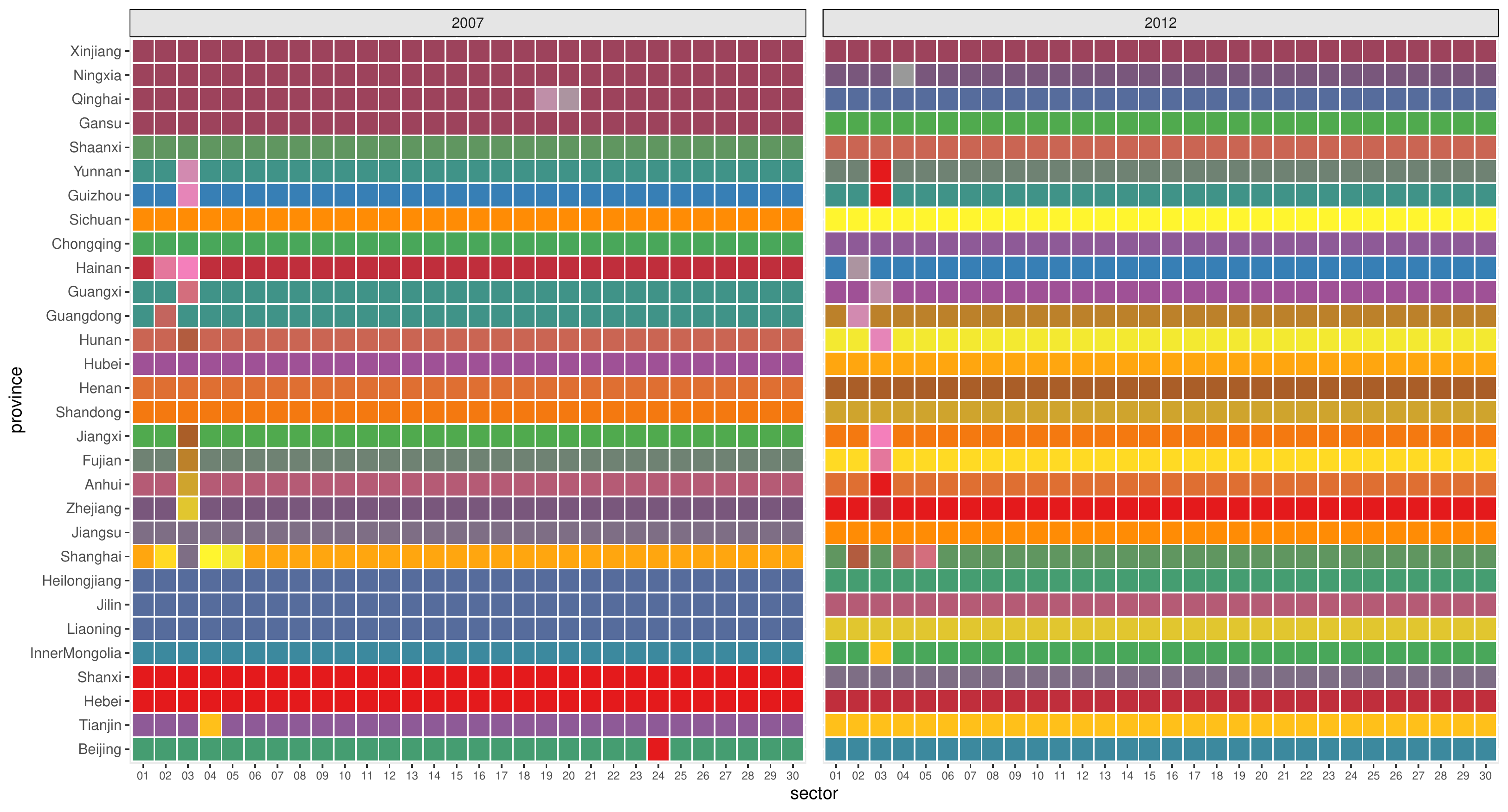}
  \caption{Communities of the MRION in 2007 (left) and 2012 (right).}
  \label{fig:cluster}
\end{figure}

The community detection results from modularity maximization are visualized in
two side-by-side heat maps in Figure~\ref{fig:cluster} for 2007 and 2012. For
each heat map, province-sectors in the same community have the same color.
Between the two years, however, the colors are not comparable because these
colors are nominal within each community detection task. There were~39 and 40
communities in 2007 and 2012, respectively. A common feature is that most
sectors from the same province belong to the same community. This is expected,
as intra-province economic ties are naturally tighter than inter-province
economic ties for geographic, historical, and administrative reasons.
An interesting discovery is that heavy industry
sectors, such as ``coal mining and processing'' (02), ``petroleum 
and gas extracting'' (03), and ``metals mining/processing'' (04) 
usually form singletons independent from province-based 
communities. For example, Shanghai as a manufacturing and business 
center usually inquires a high demand of raw materials like coal, 
which heavily relies on the supplies from other provinces. 
Consequently, ``coal mining and processing'' (02) of
Shanghai forms a singleton instead of falling into the same 
community formed by the most of the other sectors in Shanghai.

From 2007 to 2012, the community structure has shown a notable 
change.
Sectors from the provinces in the same geographic region tended to 
stay in the same
community in 2007. For example, three northeastern provinces Heilongjiang,
Jilin, and Liaoning were in one community; four northwestern 
provinces
Xinjiang, Ningxia, Qinghai, and Gansu belonged to another community; 
two central north provinces Shanxi and Hebei were placed in the same 
community. In 2012, however,
this pattern was no longer observed, each province appearing
to be a community of its own. A closer look at the data reveals that
the growth rate of
inter-province trade is much smaller than that of intra-province trade. In
2007, there were $9.6$ trillion CNY of inter-province trade and $35.5$ trillion 
CNY intra-province trade. In 2012, the inter-province trade increased to
$12.8$ trillion CNY (an increase of 30\%), while the intra-province trade 
increased to $68.4$ trillion CNY (an increase of 90\%). In some provinces, Jilin
for instance, there has been almost no change in inter-province trade from
2007 to 2012, whereas the intra-province trade has been tripled.

The community detection results again echos the regional fragmentation of the
Chinese economy at the province level \citep{poncet2005fragmented}. In the
early 2000s, under the effect of regional integration strategies such as the Great
Western Development and Northeast Revitalization, the transactions 
among the sectors
in certain regions mushroomed, forming multi-province communities. In response
to the 2008 global financial crisis, the central government advocated a
four trillion Chinese economic stimulus plan, which sustained the economic
growth while the world economy slowed down \citep{ouyang2015treatment}. The
majority of these funds were reallocated from the budget of provincial and local
governments, which supported infrastructure projects and housing
developments; some assisted local governments to lend to state-owned
companies to develop housing estates \citep{chen2013government}.
Therefore, by 2012, the extent of the regional fragmentation has
expanded despite
the goals of the regional integration strategies in the early 2000s. The extent
of the regional fragmentation in Chinese economy after 2012 is of great interest
when the 2017 MRIOT becomes available.

\subsection{Centrality}
\label{sec:res_centrality}

The PR scores were used to rank the relative importance of 
province-sectors in the
Chinese economy. The dampling factor was set to be
$\gamma = 0.85$ as recommended by~\citet{brin1998anatomy}. For different
values of $\theta = \{0, 1\}$, we computed the PR scores for 
the 900 province-sectors in 2007 and 2012 with or without 
accounting for prior information. Specifically, the total value 
added (TVA) of each province-sector was adopted as node-specific 
prior information, as it indicates the value added contributed by 
each sector to the national economy. The TVAs of province-sectors in
each year are recorded in quadrat~IV of the MRIOT; see Table~\ref{tab:mriot}. We
present the province-sectors with top~10 PR scores in Table~\ref{tab:pr}.

\begin{table}[tbp]
	\centering
	\caption{The top~10 province-sectors in PR scores ($\gamma = 0.85$)
          from the MRIONs in 2007 and 2012 with and without TVA (total
          value-added) as prior information.}
	\small \setlength{\tabcolsep}{3.7pt}
	\label{tab:pr}
	\begin{tabular}{ccccccc}
		\toprule
		\multirow{3}{*}{Rank} & \multicolumn{2}{c}{$\theta = 0$}
		& \multicolumn{4}{c}{$\theta = 1$}
		\\ \cmidrule(lr){2-3} \cmidrule(lr){4-7}
		& \multicolumn{2}{c}{No prior}
		& \multicolumn{2}{c}{No prior}
		& \multicolumn{2}{c}{Using TVA as prior}
		\\ \cmidrule(lr){2-3} \cmidrule(lr){4-5} \cmidrule(lr){6-7}
		& 2007 & 2012 & 2007 & 2012 & 2007 & 2012 \\
		\midrule
		1 & Beijing 15 & Hainan 21 & Beijing 30 & Yunnan 24 & Guangdong 30 & 
		Yunnan 24 \\ 
		2 & Beijing 10 & Hainan 12 & Guangdong 30 & Zhejiang 24 & Beijing 30 & 
		Guangdong 30 \\ 
		3 & Beijing 14 & Hainan 06 & Yunnan 24 & Guangdong 24 & Guangdong 19 & 
		Guangdong 24 \\ 
		4 & Beijing 16 & Hainan 17 & Guangdong 19 & Jiangsu 17 & Jiangsu 30 & 
		Zhejiang 24 \\ 
		5 & Beijing 12 & Hainan 14 & Guangdong 18 & Hebei 24 & Shandong 06 & 
		Jiangsu 30 \\ 
		6 & Beijing 13 & Hainan 16 & Guangdong 17 & Shaanxi 24 & Zhejiang 30 & 
		Jiangsu 17 \\ 
		7 & Beijing 17 & Hainan 09 & Jiangsu 30 & Shanghai 24 & Guangdong 18 & 
		Shandong 06 \\ 
		8 & Jiangsu 17 & Hainan 13 & Guangdong 08 & Qinghai 24 & Shanghai 30 & 
		Jiangsu 12 \\ 
		9 & Jiangsu 18 & Tianjin 12 & Shandong 06 & Sichuan 24 & Jiangsu 12 & 
		Shandong 12 \\ 
		10 & Jiangsu 22 & Hebei 06 & Shanghai 30 & Xinjiang 24 & Shandong 30 & 
		Guangdong 19 \\
		\bottomrule
	\end{tabular}
\end{table}

Without considering the weight ($\theta = 0$), most of the top~10
province-sectors are from Beijing in 2007 and from Hainan in 2012. 
Beijing's top ranking in 2007 may be explained by its special 
function as the nation's capital with an advantage in access to the 
resources nationwide. The preparation for the 2008 Olympic Games in 
urban infrastructure, ecological environment, electronic
technology and other aspects had a huge pull effect on the economic
development of Beijing, especially in the manufacturing sectors. Nonetheless,
the sizes of the sectors in Beijing are smaller than those in 
eastern coastal provinces such as
Guangdong or Jiangsu. The top ranking of Beijing without weight is, 
therefore, not consistent with wide perceptions. The top ranking of 
seven sectors in Hainan in 2012 is even more puzzling. As an island 
with a small population, Hainan is known to be relatively less 
developed compared with other provinces in China. The unweighted PR 
scores are not satisfying in measuring the centrality in the context 
of MRIONs.

With PR scores fully based on weights instead of counts ($\theta = 
1$), the top~10 province-sectors have changed dramatically. 
Province-wise, most of the top~10 province-sectors in 2007 are from
the eastern coast (Guangdong, Jiangsu, 
Shandong, and Shanghai). These provinces are more
developed than others, and consistently make large contribution to 
the national GDP. In 2012, however, some sectors from much less 
developed provinces (Shaanxi, Qinghai, and Xinjiang) joined those 
from traditional developed provinces in the top~10. This may be a 
result of these less developed provinces in northwest China 
benefiting from the effectiveness of the China Western Development 
policy. Sector-wise, the sector of ``other services'' (30) appeared 
most often in the top~10 in 2007, but ``construction'' (24) became 
dominant in 2012. Note that ``other services'' cover some essential 
services like financial services and information technology, both of 
which are critical to modern economic development. It is evident 
that these services have provided unprecedented support to the 
growth of many other sectors in China in the early 2000s. The 2008 
global financial crisis stroke many such services. Much of the 
four-trillion CNY stimulus program funded projects like railway, 
highway, bridge, and aviation construction. Further, one of the aims 
of the China Western Development program was to strengthen the 
infrastructure construction in the participating provinces.

Utilization of TVA as prior information led to
noticeable changes in both lists of top~10 PR province-sectors.
The changed results are more consistent with the intuition more than
otherwise. For 2007, Guangdong became less dominant than otherwise, albeit still
with the highest frequency in the top~10. Most of the provinces came from the 
eastern/southern coast (except for Beijing). Different from the diversity in
province, ``other services'' (30) appeared to be most influential 
sector-wise, as it occupied six positions of the top ten.
The impact of the TVA prior is more notable in 2012 than in 2007.
Sectors from western provinces like Shaanxi,
Qinghai and Xinjiang were gone, while sectors from the 
coastal provinces emerged in the top~10 list. The ``construction'' (24) sector
is less dominant, but remaining most frequent in the top~10. Other leading 
province-sectors are ``other services'' (30) from Guangdong and 
Jiangsu and ``chemical industry'' (12) from Jiangsu and Shandong.
The updated results with the TVA prior makes more sense 
because TVA contains information about the self-loops which were
otherwise discarded but are useful in assessing centrality.

It is worth special attention that ``construction'' (24) of Yunnan is 
top~1 in 2012 in spite of the inclusion of prior information. Although
a developing inland province, Yunnan is one of the largest tourist
province in China. The fast development of tourism in Yunan has
boosted the development of infrastructure construction such as
transportation facilities and hotel accommodations.
Located in the geographical center of Asia, connecting southeast 
Asia with China and inland with coastal regions, Yunnan has been 
crucial to the China Western Development. With favorable domestic 
policies and economic cooperation with Southeast Asian countries, 
Yunnan's GDP grew with a rate consistently higher than the national average
during this period, for which the construction sector played an important 
pulling role \citep{su2014multi}.

\begin{table}[tbp]
	\centering
	\caption{The province-sectors with top~10 PR scores ($\gamma = 
	0.85$,
	$\theta = 1$) of the backbones of MRIONs in 2007 and 2012 at 
	significance level $\alpha = \{10^{-3}, 10^{-4}\}$. TVA is used 
	as prior information. The blue province-sectors with * do 
	not appear in the corresponding columns for $\alpha = 1$ 
	(original networks).}
	\footnotesize \setlength{\tabcolsep}{3.7pt}
	\label{tab:pr_backbone}
	\begin{tabular}{ccccccc}
		\toprule
		\multirow{3}{*}{Rank} & \multicolumn{2}{c}{$\alpha = 1$}
		& \multicolumn{2}{c}{$\alpha = 10^{-3}$}
		& \multicolumn{2}{c}{$\alpha = 10^{-4}$}
		\\ \cmidrule(lr){2-3} \cmidrule(lr){4-5} \cmidrule(lr){6-7}
		& 2007 & 2012 & 2007 & 2012 & 2007 & 2012 \\
		\midrule
		1 & Guangdong 30 & Yunnan 24 & Guangdong 30 & Guangdong 24 & Guangdong 
		30 & Guangdong 24 \\ 
		2 & Beijing 30 & Guangdong 30 & Beijing 30 & Guangdong 30 & Beijing 30 
		& Guangdong 30 \\ 
		3 & Guangdong 19 & Guangdong 24 & Guangdong 19 & Zhejiang 24 & 
		Guangdong 19 & Jiangsu 17 \\ 
		4 & Jiangsu 30 & Zhejiang 24 & Shandong 06 & Jiangsu 17 & Shandong 06 & 
		Zhejiang 24 \\ 
		5 & Shandong 06 & Jiangsu 30 & Jiangsu 30 & Yunnan 24 & Jiangsu 30 & 
		Yunnan 24 \\ 
		6 & Zhejiang 30 & Jiangsu 17 & Guangdong 18 & Jiangsu 30 & 
		\blue{Guangdong 08*} & Jiangsu 30 \\ 
		7 & Guangdong 18 & Shandong 06 & \blue{Guangdong 08*} & 
		Shandong 06 & 
		Guangdong 18 & Shandong 06 \\ 
		8 & Shanghai 30 & Jiangsu 12 & \blue{Guangdong 17*} & 
		Jiangsu 12 & Zhejiang 30 
		& Jiangsu 12 \\ 
		9 & Jiangsu 12 & Shandong 12 & Zhejiang 30 & Guangdong 19 & 
		\blue{Guangdong 17*} & Shandong 12 \\ 
		10 & Shandong 30 & Guangdong 19 & Shandong 30 & Shandong 12 & Shandong 
		30 & Guangdong 19 \\
		\bottomrule
	\end{tabular}
\end{table}

The PR score provides another opportunity to demonstrate the effectiveness of
backbone. Table~\ref{tab:pr_backbone} summarizes the province-sectors with 
top~10 PR scores in the backbones of MRIONs with significance level 
$\alpha = \{10^{-3}, 10^{-4}\}$. No drastic change in the lists for both 2007
and 2012 are observed. Only two new province-sectors,
``clothing, leather, fur'' (08) and ``transport equipment'' 
(17) from Guangdong, ranked among the top~10 in both of the 
backbones but not in the original 2007 MRION. In fact, they were 
ranked respectively at 13 and 14 in the original MRION, with small 
difference in the magnitude of PR score from the bottom of top~10.  
For 2012, the province-sectors in the top~10 list 
remained the same for both backbones, in spite of some changes in 
order. The traditional strong sectors ``construction'' (24) and 
''other services'' (30) in Guangdong, ``transport equipment'' (17) 
in Jiangsu and ``construction'' (24) in Zhejiang surpassed 
``construction'' (24) in Yunnan. This was because 
``construction'' (24) in Yunnan is more connected by insignificant 
edges in the MRION of 2012, but these edges are filtered out in the 
backbones. Between the two lists for the backbones at 
different significance levels, we only observe small dispersion in 
province-sector orders. For instance, in the two lists of 2012, the 
province-sectors at rank~3 and~4 and those at rank 9 and 10 were 
respectively switched, with the rest remaining identical. This comparison,
again, shows that backbone is capable of capturing the centrality measures of
the vertices in the MRIONs.

\section{Discussions}
\label{sec:dis}

MRIOTs provide a natural arena for network analyses to study regional and
sectoral structure of an economy.  Our study of MRIONs of China in 2007 and 2012
is the first network analysis of its kind for the Chinese economy.
All three types of strength distributions (in, out, and total) were found to be
skewed to the left, where the degree of skewness of each increases over time.
For each MRION, the positive assortative coefficients suggests assortative
mixing across the province-sectors, especially intra-province-sectors.
As indicated by close-to-one clustering coefficients, the province-sectors tend
to cluster together, and the tendency increased inter-province transactions. 
Province-based community structures were detected. There were communities
containing multiple provinces In 2007 but none in 2012, suggesting increased
regional fragmentation. The most essential province-sectors in the Chinese
economy were found through a new class of weighted PR measures. Province-wise,
Guangdong, Jiangsu and many eastern coastal provinces contain most 
sectors in the top list. Sector-wise,  ``construction'' (24) and ``other
services'' (30) appear to be dominant.

Our study suggests a few methodological caveats for network analysis of MRIONs.
First, it is critical to account for edge weight when summarizing MRIONs as
these networks are weighted; otherwise, misleading inference may be made. For
example, unweighted assortativity coefficients suggest almost no 
assortative (or disassortative) mixing in the MRIONs, whereas the 
weighted counterparts indicate a moderately positive assortative 
mixing. While no weight is accounted, the classical PR algorithm 
have produced a top~10 list in 2012 containing 8 sectors form 
Hainan, which is not consistent with the fact. Rather, the weighted 
PR algorithm (with or without using TVA as prior information) has 
provided a much more reasonable result. Second, precise 
interpretation of the network measures is crucial. For instance,
clustering coefficient is a relative measure. A large local clustering
coefficient of a vertex only tells how likely its two neighbors are connected;
it says nothing about how many neighbors it has. For instance, the 
province-sector of the highest local clustering coefficient in 2012 
is given to ``construction'' (24) from Inner Mongolia according to 
the computation. This specific province-sector receives a high value 
since it has fewer neighbors compared to the rest in the nation, 
leading to a higher proportion conversely. Lastly, when ranking the 
vertices in a network, we recommend to make full use of the possible 
vertex-specific auxiliary information, which helps lead to more 
intuitive results.
These caveats may be applicable to networks beyond MRIONs.

\section*{Acknowledgments}

Tao Wang and Shiying Xiao were supported by the National Bureau of 
Statistics of China (2018lz33).

\appendix

\section{Province codes}
\label{sec:provcode}

Table~\ref{tab:prov} summarizes the code and the names of the 30 provinces used
in the chord graph in Figure~\ref{fig:chord}.

\begin{table}[tbp]
	\centering
	\caption{Codes of the provinces in the MRIONs}
	\label{tab:prov}
	\begin{tabular}{cccc}
		\toprule
		Code & Province & Code & Province \\
		\midrule
		01 & Beijing & 16 & Henan \\ 
		02 & Tianjin & 17 & Hubei \\ 
		03 & Hebei & 18 & Hunan \\ 
		04 & Shanxi & 19 & Guangdong \\ 
		05 & Inner Mongolia & 20 & Guangxi \\ 
		06 & Liaoning & 21 & Hainan \\ 
		07 & Jilin & 22 & Chongqing \\ 
		08 & Heilongjiang & 23 & Sichuan \\ 
		09 & Shanghai & 24 & Guizhou \\ 
		10 & Jiangsu & 25 & Yunnan \\ 
		11 & Zhejiang & 26 & Shaanxi \\ 
		12 & Anhui & 27 & Gansu \\ 
		13 & Fujian & 28 & Qinghai \\ 
		14 & Jiangxi & 29 & Ningxia \\ 
		15 & Shandong & 30 & Xinjiang \\
		\bottomrule
	\end{tabular}
\end{table}

\section{Four types of local clustering coefficients}
\label{sec:triangle}

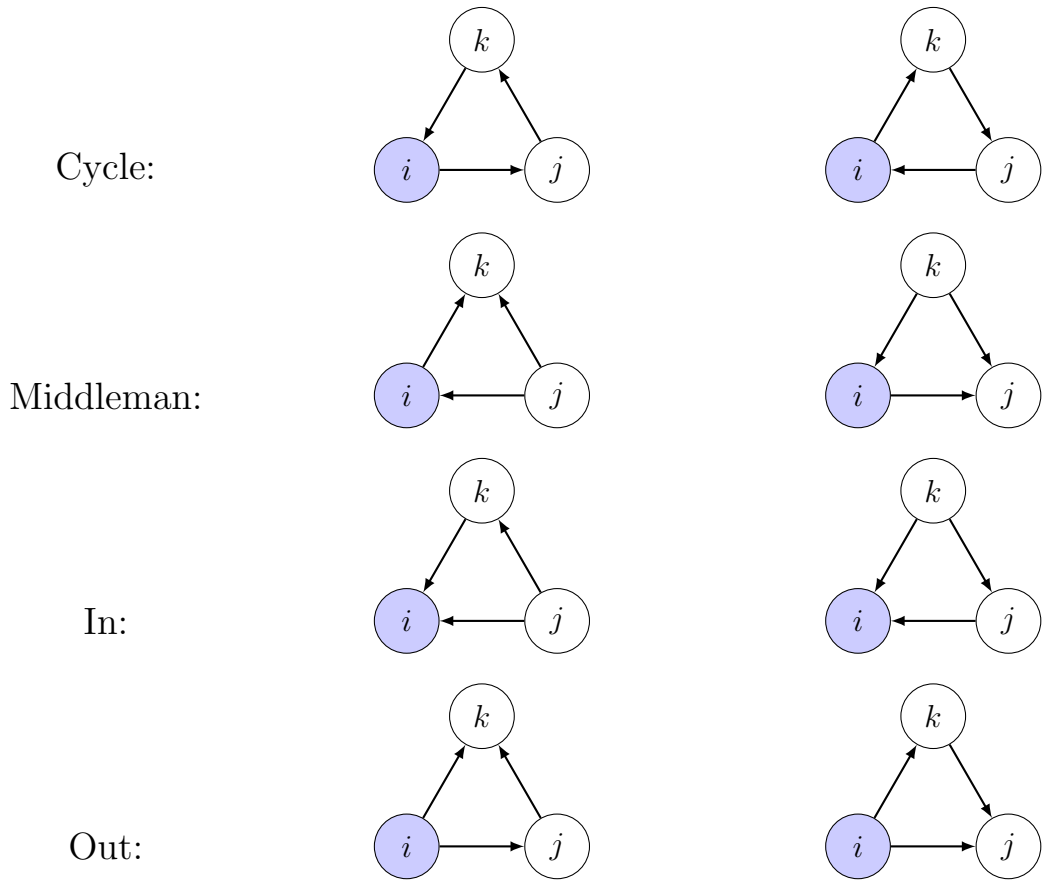
\begin{figure}[tbp]
	\begin{center}
		\begin{tikzpicture}
			\draw (-4,0) node {\rm \large Cycle:} ;
			\draw (0,0) node[draw = black, circle, minimum size =
			0.86cm, fill = blue!20] (cyci1)	{$i$} ;
			\draw (2,0) node[draw = black, circle, minimum size =
			0.86cm,] (cycj1) {$j$} ;
			\draw (1,1.732) node[draw = black, circle, minimum size
			= 0.86cm,] (cyck1) {$k$} ;
			\draw[-latex, thick] (cyci1) -- (cycj1) ;
			\draw[-latex, thick] (cycj1) -- (cyck1) ;
			\draw[-latex, thick] (cyck1) -- (cyci1) ;
			\draw (6,0) node[draw = black, circle, minimum size =
			0.86cm, fill = blue!20] (cyci2)	{$i$} ;
			\draw (8,0) node[draw = black, circle, minimum size =
			0.86cm,] (cycj2) {$j$} ;
			\draw (7,1.732) node[draw = black, circle, minimum size
			= 0.86cm,] (cyck2) {$k$} ;
			\draw[-latex, thick] (cyci2) -- (cyck2) ;
			\draw[-latex, thick] (cyck2) -- (cycj2) ;
			\draw[-latex, thick] (cycj2) -- (cyci2) ;
			\draw (-4,-3) node {\rm \large Middleman:} ;
			\draw (0,-3) node[draw = black, circle, minimum size =
			0.86cm, fill = blue!20] (midi1)	{$i$} ;
			\draw (2,-3) node[draw = black, circle, minimum size =
			0.86cm,] (midj1) {$j$} ;
			\draw (1,-1.268) node[draw = black, circle, minimum size
			= 0.86cm,] (midk1) {$k$} ;
			\draw[-latex, thick] (midi1) -- (midk1) ;
			\draw[-latex, thick] (midj1) -- (midi1) ;
			\draw[-latex, thick] (midj1) -- (midk1) ;
			\draw (6,-3) node[draw = black, circle, minimum size =
			0.86cm, fill = blue!20] (midi2)	{$i$} ;
			\draw (8,-3) node[draw = black, circle, minimum size =
			0.86cm,] (midj2) {$j$} ;
			\draw (7,-1.268) node[draw = black, circle, minimum size
			= 0.86cm,] (midk2) {$k$} ;
			\draw[-latex, thick] (midk2) -- (midi2) ;
			\draw[-latex, thick] (midi2) -- (midj2) ;
			\draw[-latex, thick] (midk2) -- (midj2) ;
			\draw (-4,-6) node {\rm \large In:} ;
			\draw (0,-6) node[draw = black, circle, minimum size =
			0.86cm, fill = blue!20] (ini1)	{$i$} ;
			\draw (2,-6) node[draw = black, circle, minimum size =
			0.86cm,] (inj1) {$j$} ;
			\draw (1,-4.268) node[draw = black, circle, minimum size
			= 0.86cm,] (ink1) {$k$} ;
			\draw[-latex, thick] (inj1) -- (ini1) ;
			\draw[-latex, thick] (ink1) -- (ini1) ;
			\draw[-latex, thick] (inj1) -- (ink1) ;
			\draw (6,-6) node[draw = black, circle, minimum size =
			0.86cm, fill = blue!20] (ini2)	{$i$} ;
			\draw (8,-6) node[draw = black, circle, minimum size =
			0.86cm,] (inj2) {$j$} ;
			\draw (7,-4.268) node[draw = black, circle, minimum size
			= 0.86cm,] (ink2) {$k$} ;
			\draw[-latex, thick] (inj2) -- (ini2) ;
			\draw[-latex, thick] (ink2) -- (ini2) ;
			\draw[-latex, thick] (ink2) -- (inj2) ;
			\draw (-4,-9) node {\rm \large Out:} ;
			\draw (0,-9) node[draw = black, circle, minimum size =
			0.86cm, fill = blue!20] (outi1)	{$i$} ;
			\draw (2,-9) node[draw = black, circle, minimum size =
			0.86cm,] (outj1) {$j$} ;
			\draw (1,-7.268) node[draw = black, circle, minimum size
			= 0.86cm,] (outk1) {$k$} ;
			\draw[-latex, thick] (outi1) -- (outj1) ;
			\draw[-latex, thick] (outi1) -- (outk1) ;
			\draw[-latex, thick] (outj1) -- (outk1) ;
			\draw (6,-9) node[draw = black, circle, minimum size =
			0.86cm, fill = blue!20] (outi2)	{$i$} ;
			\draw (8,-9) node[draw = black, circle, minimum size =
			0.86cm,] (outj2) {$j$} ;
			\draw (7,-7.268) node[draw = black, circle, minimum size
			= 0.86cm,] (outk2) {$k$} ;
			\draw[-latex, thick] (outi2) -- (outj2) ;
			\draw[-latex, thick] (outi2) -- (outk2) ;
			\draw[-latex, thick] (outk2) -- (outj2) ;
		\end{tikzpicture}
	\end{center}
\caption{Four types of triangles proposed
in \citet{fagiolo2007clustering}, and reused
in \citet{clemente2018directed}.}
\label{fig:triangles}
\end{figure}

In practice, the formulations of the four kinds of local clustering
coefficients are respectively given by
\begin{align}
	\label{eq:clust_in}
	C^{\, \rm in}_i &= \frac{\left[\wei^{\top} \left(\adj +
		\adj^{\top}\right) \adj\right]_{ii}}{2\stin_i
		\left(\din_i - 1\right)},
	\\ \label{eq:clust_out}
	C^{\, \rm out}_i &= \frac{\left[\wei \left(\adj +
		\adj^{\top}\right) \adj^{\top}\right]_{ii}}{2\stout_i
		\left(\dout_i - 1\right)},
	\\ \label{eq:clust_mid}
	C^{\, \rm mid}_i &= \frac{\left(\wei^{\top} \adj
		\adj^{\top} +
		\wei \adj^{\top} \adj\right)_{ii}}{\left(\stin_i
		\dout_i +
		\stout_i \din_i \right) - \left(\adj \wei +
		\wei \adj\right)_{ii}},
	\\ \label{eq:clust_cyc}
	C^{\, \rm cyc}_i &= \frac{\left[\wei \adj^2 +
		\wei^{\top}\left(\adj^{\top}\right)^2\right]_{ii}}{\left(
		\stin_i \dout_i + \stout_i \din_i \right) - \left(\adj
		\wei
		+
		\wei \adj\right)_{ii}}.
\end{align}

\end{document}